\documentclass[a4paper,onecolumn,11pt,accepted=2025-06-17]{quantumarticle}
\pdfoutput=1
\usepackage[utf8]{inputenc}
\usepackage[english]{babel}
\usepackage[T1]{fontenc}
\usepackage[numbers]{natbib}
\usepackage{listings}

\usepackage{amsmath}
\usepackage{amssymb}
\usepackage{hyperref}

\usepackage{latexsym}
\usepackage{amsmath, a4wide, amssymb, amsfonts,  epsfig, xcolor}
\usepackage{mathrsfs}
\usepackage{url}
\usepackage{graphicx}
\usepackage{caption}
\usepackage{subcaption}

\usepackage{placeins}

\makeatletter
\@addtoreset{equation}{section}
\makeatother

\newcommand{\diag}{{\rm diag}}
\newcommand{\vecc}{{\rm vec}}
\newcommand{\xvec}{\mathbf{x}}

\newcommand{\qstate}[1]{\mid \! {#1} \rangle}
\newcommand{\qqstate}[1]{ \langle  {#1} \!\mid}
\newcommand{\sqabs}[1]{ \left| #1 \right| ^2}
\newcommand{\RE}{\mathrm{Re}}
\newcommand{\IM}{\mathrm{Im}}
\newcommand{\expbig}[1]{\exp{\big( {#1} \big)}}
\newcommand{\expBig}[1]{\exp{\Big( {#1} \Big)}}

\newcommand{\Hhat}{\hat{\mathbf{H}}}

\newcommand{\Vhat}{\hat{\mathbf{V}}}

\newcommand{\II}{\mathbf{I}}
\newcommand{\IB}{\mathbf{I}}
\newcommand{\AAA}{\mathbf{A}}
\newcommand{\BB}{\mathbf{B}}
\newcommand{\CB}{\mathbf{C}}
\newcommand{\HH}{\mathbf{H}}
\newcommand{\MM}{\mathbf{M}}
\newcommand{\SB}{\mathbf{S}}
\newcommand{\UU}{\mathbf{U}}
\newcommand{\UD}{\underline{\UU}}

\newcommand{\KK}{\mathbf{K}}

\title{Experiments with Schr{\" o}dinger Cellular Automata}
\author{Kees van Berkel}
\email{c.h.v.berkel@tue.nl}
\orcid{https://orcid.org/0000-0002-3577-575X}
\author{Jan de Graaf}
\thanks{passed away on 2024, September 16.}
\author{Kees van Hee}
\email{k.m.v.hee@tue.nl}
\affiliation{TU Eindhoven, Dept. of Mathematics \& Computer Science,
 	P.O. Box 513, 5600 MB Eindhoven, The Netherlands}

\begin{document}

\maketitle

\begin{abstract} 
We derive a class of cellular automata for the Schr{\"o}dinger Hamiltonian,
including scalar and vector potentials.
It is based on a multi-split of the Hamiltonian,
resulting in a multi-step unitary evolution operator in discrete time and space.
Experiments with one-dimensional automata offer quantitative insight
in phase and group velocities, energy levels, related approximation errors,
and the evolution of a time-dependent harmonic oscilator.
The apparent effects of spatial waveform aliasing are intriguing.
Interference experiments with two-dimensional automata include refraction,
 Davisson-Germer, Mach-Zehnder, single \& double slit, and Aharonov-Bohm.
\end{abstract}


\section{Introduction}
\label{sec:introduction}

The Schr{\" o}dinger equation is a linear partial differential equation 
that governs the non-relativistic evolution of the wave function of a quantum-mechanical system.
For a single particle it has the form
\cite{2009-Feynman-Vol3}, Vol.\ III, 21-1
\begin{equation}
\begin{array}{lll}
	i \hbar \dfrac{\partial \Psi(\xvec,t)}{\partial t} 
    			&=&  {\cal H}\; \Psi(\xvec,t) \\[2mm]
    	{\cal H} 	&=&   \dfrac{1}{2m} 
    		\left(\dfrac{\hbar}{i} \nabla - q \AAA(\xvec) \right) \cdot
		\left(\dfrac{\hbar}{i} \nabla - q \AAA(\xvec) \right) 
		+ q \phi(\xvec)  ~,
\end{array}
\label{eq:schroedinger}
\end{equation} 
where $\Psi(\xvec,t)$ is the three-dimensional wave function,  
${\cal H}$ is the so-called Hamiltonian (operator), 
and $\hbar$ the Planck constant.
In addition, $m$ denotes the particle's mass, $q$ the particle's charge,
$\AAA(\xvec)$ the vector potential and $\phi(\xvec)$ the scalar potential.
The Hamiltonian is time-independent since neither $\phi(\xvec)$ nor $\AAA(\xvec)$ depend on time.
Usually the vector potential $\AAA(\xvec)$ is not considered, resulting in a simpler version of the Hamiltonian.
\begin{equation}
	{\cal H} = - \frac{\hbar^2}{2m} \nabla^2  +V(\xvec) ~,
\label{eq:schroedinger-V}
\end{equation} 
where $V(\xvec) = q \phi(\xvec)$ is the potential(-energy) function. 
For a free particle, $V(\xvec) = 0$ and the Hamiltonian further reduces to 
\begin{equation}
	{\cal H} = - \frac{\hbar^2}{2m} \nabla^2  ~.
\label{eq:schroedinger-free}
\end{equation} 

Textbooks such as  \cite{2018-Griffiths} present analytical solutions for a variety of simple quantum systems.
For systems with non-trivial $V(\xvec)$ or $\AAA(\xvec)$ numerical techniques can be used 
to approximate the continuous solution of $\Psi$ in discrete time and space.
This paper explores the opposite perspective: 
what if the Schr{\" o}dinger equation is a continuous approximation of a discrete universe?
{\em What if}, e.g. at the Planck scale, quantum dynamics occurs on a discrete lattice and in discrete time steps, 
as for example in \cite{1990-thooft, 2014-thooft, 2011-dAriano, 2014-Arrighi, 2015-bisio}?
For this exploration cellular automata are used as a tool.
Quoting \cite{1987-Toffoli}:
\begin{quote}
  Cellular automata are discrete dynamical systems whose behavior is completely specified in terms of a local relation, 
  much as is the case for a large class of continuous dynamical systems defined by partial differential equations. 
  {\em In this sense, cellular automata are the computer scientist's counterpart to the physicist's concept of ``field.''}
\end{quote}
The Schr{\" o}dinger cellular automata (SCA) consist of a finite number, $N$, of cells,
where the state of each cell is a complex number representing the value of the wave function at that location.
The automata are linear and the update rule (``local relation''  in the quote above) 
can therefore be represented by an $N \!\times\! N$ evolution matrix $\UU$. 
Matrix $\mathbf{U}$ must be both {\em unitary}, to preserve Born probability, 
and {\em local} to meet the locality requirement of cellular automata.
For a 1D SCA this locality means that $\UU$ must be band-structured.

\subsection{Contributions}
\label{subsec:contributions}

The contributions of this work include the following.
\begin{enumerate}
\item A systematic derivation of an evolution matrix $\mathbf{U}$ 
	for the 3D Schr{\" o}dinger Hamiltonian (\ref{eq:schroedinger}) that is both local and unitary,
	with ${\mathcal O}(\tau^3)$ error for time step $\tau$.
	(Section \ref{sec:Schroedinger-CA}.)
\item A derivation of the phase velocity, group velocity, and dispersion relation,
	as  well as the energy levels of the infinite potential well and the harmonic oscillator.
	These include analytic expressions for the $\mathbf{U}$-approximation error,
	and descriptions of aliasing effects resulting from space discretization.
        (Section \ref{sec:1D-experiments}.)
\item Experiments with SCA describing a time-dependent harmonic oscillator, 
         including adiabatic heating (Subsection \ref{subsec:adiabiatic}), 
	2D reflection and refraction, the Davisson-Germer experiment,
	the Mach-Zehnder interferometer, the single and double-slit experiments,
	and two variants of the Aharonov-Bohm effect.
	(Section \ref{sec:2D-experiments}.)
\end{enumerate}
The Python code for the SCA experiments is available on github \cite{2025-vBerkel-github}
and the animations can be viewed on youtube \cite{2025-vBerkel-youtube}.

\subsection{Related work}
\label{subsec:related}

 {\bf Numerical techniques} can be applied to solve the Schr{\" o}dinger equation,
 typically based on the  Crank-Nicolson method \cite{1947-Crank}.
Mena \cite{2023:Mena}, amongst others, applied this method to various systems, 
including the double-slit experiment based a 2D grid of $N=160 \times160$ cells.
The evolution matrix $\mathbf{U}$ combines  forward Euler and backward Euler integration,
and is therefore dense, with all its elements nonzero. 
This non-locality is incompatible with cellular automata and also implies a very high computational load.

{\bf Schr{\" o}dinger CA} were pioneered by Gr\"{o}ssing and Zeilinger \cite{1988-Grossing}.
However, their evolution matrix $\mathbf{U}$ is not unitary,
which introduces a need for periodic normalization.
Meyer \cite{1996-Meyer} identified a class of cellular automata where $\mathbf{U}$ 
is both unitary and band structured, based on the work of Toffoli  \cite{1987-Toffoli}
on so-called {\em partitioned cellular automata}.
Boghosian and Taylor \cite{1998-Boghosian} propose a first SCA and showed that 
in the continuum limit it obeys the Schr{\" o}dinger equation. 
In Section \ref{subsec:split-hamiltonian} we derive essentially the same SCA. 
They also discuss the addition of an external (scalar) potential and offer a dispersion relation.

 {\bf Dirac cellular automata} (DCA) are cellular automata based on the Dirac equation, as proposed by Bialynicki-Birula \cite{1994-Bialynicki}.
D'Ariano \cite{2011-dAriano} extended this work, where he described matrix $\mathbf{U}$ as a quantum circuit.
This circuits ``consist of alternate uniform rows of gates'' inspired by partitioned CA.
Bisio et al \cite{2013-bisio} derive the phase/group velocities for DCA,
not addressing periodicity in $k$ or aliasing.
See also Costa \cite{2021-Costa} for background 
and Arrighi \cite{2019-Arrighi} and Farrelly \cite{2020-Farrelly} for review articles on 
the broad area of quantum cellular automata.

 {\bf Quantum walks} (QW) is an extensive field of research related to quantum cellular automata
and was pioneered by Aharonov et al \cite{1993-Aharonov}.
These quantum walks can also be modelled by a unitary matrix $\mathbf{U}$,
where the state and evolution are represented in a different way, based on the notion of a ``quantum coin".
Ambainis et al \cite{2005-Ambainis} describe a specific class of quantum walks as
``an alternation of coin flip and moving step'',  adopting the two-step CA approach of \cite{1996-Meyer}.
Both alternate QWs (AQW, Bru et al \cite{Bru-2016}),
and staggered QWs (Portugal et al \cite{2015-Portugal}) are based on this two-step approach,
with Strauch \cite{2006-Strauch} describing a first Schr{\" o}dinger  QW.
Generalization to 3+1 dimensions are addressed by Arrighi et al \cite{2014-Arrighi} and by
D'Ariano et al \cite{2016-Dariano}.
Costa et al \cite{2018-Costa} offer a translation of staggered QW to partitioned unitary quantum CAs (PUQCAs).
Jolly et al \cite{2023-Jolly} introduce {\em twisted} PQWs (Plastic Quantum Walkers), 
which adds a dispersion term ("twist") to a class SQW/PQW that admits a continuum limit ("plasticity").
Our evolution operator $\mathbf{U}$ is explicitly dependent on $\tau$ and has the continuum limit built-in. 
Subsection \ref{subsec:qw} describes the relation between Schr{\" o}dinger QWs and Schr{\" o}dinger CAs.

Adding {\bf vector potential} $\AAA(\xvec)$ to the Hamiltonian has been explored for quantum walks. 
For the Dirac Hamiltonian this includes work by 
Di Molfetta et al \cite{2014-Di-Molfetta} in 1 spatial dimension, 
Arnault et al \cite{2016-Arnault} for a constant and uniform magnetic field, 
and Martin et al \cite{2018-Martin} in 2 spatial dimensions. 
Cedzich et al \cite{2019-Cedzich} offer a general framework vector potentials in QWs 
and apply it to the Hadamard coin. 
This work adds the vector potential to SCAs in three spatial dimensions, 
and applies the result to two variants of the Aharonov-Bohm experiment, 
viz. Mach-Zehnder and double-slit.
 
{\bf Approximation} of some form is needed to compute  $\exp{\left( i (\mathbf{A}+\mathbf{B}) \Delta t \right)}$
in the case that matrices $\mathbf{A}$ and $\mathbf{B}$ do not commute.
This unavoidably introduces an approximation error, 
which commonly is measured as the spectral norm error \cite{1999-Sornborger}.
Morales et al \cite{2024-Morales} propose an eigenvalue error, 
related to the eigenvalues of the underlying Hamiltonian, 
which allows a clear interpretation for long-time evolutions.
In Section \ref{sec:1D-experiments} the approximation error 
is measured in terms of phase/group velocities and energy levels.

{\bf Experiments} with Schr{\" o}dinger or Dirac  versions of CAs or QWs are comparatively rare
compared to the extensive work on formalization and mathematical foundations.
Furthermore, these experiments are typically limited to one dimension, as e.g. in \cite{2013-bisio}.
A notable exceptions is the 3D Dirac QW of \cite{2016-Dariano}.
Oliveira et al \cite{2007-Oliveira} describe both single-slit and double-slit experiments based on QWs,
where they apply Hadamard, Fourier, and Grover coins,
without reference to the Schr{\" o}dinger or Dirac equations.

\section{Schr{\" o}dinger Cellular Automata}
\label{sec:Schroedinger-CA}

This section describes the construction of Schr{\" o}dinger cellular automata in a number of steps:
1) derive a discrete version of the 1D Schr{\" o}dinger Hamiltonian for a free particle,
2) derive a unitary evolution matrix that is band structured by splitting the 1D Hamiltonian,
3) generalize to 3D for the free particle
4) include the scalar potential $V(\xvec)$ and vector potential  $\AAA(\xvec)$, and
5) include reflector cells to support boundary conditions.

\subsection{A discrete Schr{\" o}dinger Hamiltonian for a free particle}
\label{subsec:discrete-free}

For a one-dimensional cellular automaton with cell size $a$,
a discrete-space version of the Schr{\" o}dinger Hamiltonian (\ref{eq:schroedinger-free}) becomes\begin{equation}
    \mathbf{H} \Psi(x,t)
= - \frac{\hbar^2}{2m} \frac1{a^2} \big( \Psi(x\!-\!1,t) - 2 \Psi(x,t) + \Psi(x\!+\!1,t)   \big)  ~,
\label{eq:schroedinger-discrete}
\end{equation} 
where spatial coordinate $x$ is now integer valued. 
Hamiltonian $\mathbf{H}$, assuming periodic boundary conditions,
can be represented by a two-dimensional circulant matrix.
For $N\!=\!8$ cells 
\begin{equation}
  \begin{array}{lll}
  \mathbf{H} &=& \delta_m   \hat{\mathbf{H}} ~,		\\[10mm]
  \delta_m    &=& \dfrac{\hbar^2}{2m} \dfrac1{a^2}~,	
  \end{array} \hspace{2cm}
  \hat{\mathbf{H}} =
  \left[\begin{matrix}2 & -1 & 0 & 0 & 0 & 0 & 0 & -1\\-1 & 2 & -1 & 0 & 0 & 0 & 0 & 0\\0 & -1 & 2 & -1 & 0 & 0 & 0 & 0\\0 & 0 & -1 & 2 & -1 & 0 & 0 & 0\\0 & 0 & 0 & -1 & 2 & -1 & 0 & 0\\0 & 0 & 0 & 0 & -1 & 2 & -1 & 0\\0 & 0 & 0 & 0 & 0 & -1 & 2 & -1\\-1 & 0 & 0 & 0 & 0 & 0 & -1 & 2\end{matrix}\right] .
\label{eq:H-matrix}
\end{equation} 
The time evolution of a system is commonly described by \footnote{
 	Henceforth, $\UD$ denotes a dense evolution matrix and $\UU$ an ``SCA-sparse'' one.}
\begin{equation}
    \qstate{\Psi(t)} \;=\;  \UD(t)  \qstate{\Psi(0)} ~,   ~~~~~ \mathrm{with} ~~~~~ 
    \UD(t) = \expBig{-i \frac1\hbar \mathbf{H} t} ~.
\label{eq:evolution_step}
\end{equation} 
Matrix $\UD$ is a dense matrix: all its elements are nonzero.
The discrete-time evolution for {\em integer} time $t$, $0 \leq t$, and fixed time step $\tau$ thus becomes
\begin{equation}
     \qstate{\Psi((t+1)\tau)} \;=\;  \UD  \qstate{\Psi(t \tau)}  ~,
\label{eq:evolution_U}
\end{equation} 
where
\begin{equation} 
  	\UD 	= \expBig{-i \frac\tau\hbar \mathbf{H} }
		= \expbig{-i \theta \Hhat }~,  ~~~~~ \mathrm{with} ~~~~~ 
                    \theta = \frac\tau\hbar \delta_m =  \frac{\hbar}{2m} \frac\tau{a^2} ~.
\label{eq:schroedinger-evolution}
\end{equation} 
Constant $\theta$ is dimensionless, and can be seen as a rotation angle.
The probability of finding the particle in cell $x$ at time $t$, denoted by $P(x,t)$,
is given by Born's rule:
\begin{equation}
   P(x,t) \;=\; \sqabs{\Psi(x,t)}~,
  \hspace{3cm} P(t) = \sum_x P(x,t) = 1 ~.
\label{eq:schroedinger-P}
\end{equation} 

Gr{\" o}ssing and Zeilinger \cite{1988-Grossing} base their SCA on Hamiltonian $\delta_m(\Hhat-2\II)$
and approximate the evolution operator by using only the first-order term of its expansion
\begin{equation}      
		\UU
  	 ~=~ \mathbf{I} \!-\! i \theta  (\Hhat-2\II)  
	~~ \approx~~	\expBig{- i \theta (\Hhat-2\II)} ~.
\label{eq:U-Grossing}
\end{equation} 
Matrix $\mathbf{U}$ is a three-band matrix, which nicely agrees with the locality requirement
for a cellular automaton: the value of $\Psi(x, t\!+\!1)$ only depends on $\Psi(u, t)$, 
with $u$ an element of the local neighborhood of $x$, that is, $u \in \{x\!-\!1, x, x\!+\!1\}$.
This locality comes at a price: matrix $\mathbf{U}$ is {\em not} unitary, 
and the overall probability $P(t)$ is therefore not preserved.
In practice, this implies the need for an ad-hoc scale factor and regular normalization.

\subsection{Split Hamiltonian }
\label{subsec:split-hamiltonian}

Next we derive an evolution matrix $\UU$ that is {\em both} unitary 
(as required by quantum mechanics), and band structured (as required for cellular automata).
As a first step, Hamiltonian $\Hhat$ is split into two 2$\times$2-block diagonal matrices 
$\Hhat_0$ and $\Hhat_1$, such that $\Hhat = \Hhat_0 +\Hhat_1$, where
\begin{equation} 
   \Hhat_0 = \mathbf{I}_M \otimes \BB 					~, \hspace{2cm}
   \Hhat_1 = \SB^{-1} \Hhat_0 \;\SB    					~, \hspace{2cm}
   \BB=  \left[ \begin{matrix} 1 & -1 \\ -1 & 1 \end{matrix}\right] 	~.
\label{eq:H-split}
\end{equation} 
Here $\otimes$ denotes the Kronecker matrix product, integer $M$ satisfies $2M\!=\!N$,
and matrix $\mathbf{S}$ is the so-called {\em circular shift} matrix.
The product $\mathbf{S}$ $\mathbf{A}$ yields a specific permutation of matrix $\mathbf{A}$,
with all rows of $\mathbf{A}$ shifted down by one row and with the last row moved to the first position.
Below the result of this split for $N\!=\!8$ cells.
\begin{equation} 
\Hhat_0\!=\!\! \left[\begin{matrix}1 & -1 & 0 & 0 & 0 & 0 & 0 & 0\\-1 & 1 & 0 & 0 & 0 & 0 & 0 & 0\\0 & 0 & 1 & -1 & 0 & 0 & 0 & 0\\0 & 0 & -1 & 1 & 0 & 0 & 0 & 0\\0 & 0 & 0 & 0 & 1 & -1 & 0 & 0\\0 & 0 & 0 & 0 & -1 & 1 & 0 & 0\\0 & 0 & 0 & 0 & 0 & 0 & 1 & -1\\0 & 0 & 0 & 0 & 0 & 0 & -1 & 1\end{matrix}\right]
\Hhat_1\!=\!\! \left[\begin{matrix}1 & 0 & 0 & 0 & 0 & 0 & 0 & -1\\0 & 1 & -1 & 0 & 0 & 0 & 0 & 0\\0 & -1 & 1 & 0 & 0 & 0 & 0 & 0\\0 & 0 & 0 & 1 & -1 & 0 & 0 & 0\\0 & 0 & 0 & -1 & 1 & 0 & 0 & 0\\0 & 0 & 0 & 0 & 0 & 1 & -1 & 0\\0 & 0 & 0 & 0 & 0 & -1 & 1 & 0\\-1 & 0 & 0 & 0 & 0 & 0 & 0 & 1\end{matrix}\right].
\label{eq:H0-H1}
\end{equation} 
As matrices $\Hhat_0$ and $\Hhat_1$ do not commute, 
the evolution operator can be approximated 
\begin{equation} 
\begin{array}{lllll}
 	\expbig{-i \theta \Hhat}
  			&=& \expbig{-i \theta (\Hhat_0 + \Hhat_1)} 	\\
  			&=& \expbig{-i \theta \Hhat_1} \expbig{-i \theta \Hhat_0}  
				~+ \mathcal{O}(\theta^2) 				\\ [1mm]
			&=& \UU	~+ \mathcal{O}(\theta^2)	\\ [1mm]
	\UU		&=& \UU_1 \UU_0						\\ [1mm]
	\UU_0 	&=& \expbig{-i \theta \Hhat_0}   
				~=~ \II_M \otimes \CB ~ 				\\ [1mm]
	\UU_1 	&=& \expbig{-i \theta \Hhat_1}   
				~=~  \SB^{-1} \UU_0\;\SB ~ 			\\ [1mm]
	\CB	&=& \expbig{-i \theta \BB}
	 			~~~=~  \expbig{-i\theta} 
				\left[\begin{matrix}
					\cos{\left(\theta \right)}  & i \sin{\left(\theta \right)} \\
					i \sin{\left(\theta \right)} &  \cos{\left(\theta \right)}
				\end{matrix}\right] ~.
\end{array}
\label{eq:U_homogeneous}
\end{equation} 
Matrices $\mathbf{U}_0$ and $\mathbf{U}_1$ are both unitary and have a three-diagonal structure.\footnote{
 	Matrix $\mathbf{C}  =  \expbig{-i \theta } R_x(-2\theta)$, where $R_x(-2\theta)$ is the qubit rotation operator.}  
For $N\!=\!8$  
\begin{equation} 
\begin{array}{ll}
\mathbf{U}_1 = \expbig{-i\theta} \times \\ \vspace{1mm} ~~~~~~ 
\left[\begin{matrix}\cos{\left(\theta \right)} & 0 & 0 & 0 & 0 & 0 & 0 & i \sin{\left(\theta \right)}\\0 & \cos{\left(\theta \right)} & i \sin{\left(\theta \right)} & 0 & 0 & 0 & 0 & 0\\0 & i \sin{\left(\theta \right)} & \cos{\left(\theta \right)} & 0 & 0 & 0 & 0 & 0\\0 & 0 & 0 & \cos{\left(\theta \right)} & i \sin{\left(\theta \right)} & 0 & 0 & 0\\0 & 0 & 0 & i \sin{\left(\theta \right)} & \cos{\left(\theta \right)} & 0 & 0 & 0\\0 & 0 & 0 & 0 & 0 & \cos{\left(\theta \right)} & i \sin{\left(\theta \right)} & 0\\0 & 0 & 0 & 0 & 0 & i \sin{\left(\theta \right)} & \cos{\left(\theta \right)} & 0\\i \sin{\left(\theta \right)} & 0 & 0 & 0 & 0 & 0 & 0 & \cos{\left(\theta \right)}\end{matrix}\right]
\end{array}
 \label{eq:U1}
\end{equation} 
Matrix $\UU$ is unitary and has a five-diagonal structure.\footnote{
  Matrix $\UU$ has eigenvalue $\lambda \!=\!1$ with eigenvector $[1, 1, \ldots, 1]^\top$;
   each row sum (column sum) equals $1$.
}
For $N\!=\!8$
\begin{equation} 
\begin{array}{ll}
\UU = \expbig{-2i\theta} \times \\ \vspace{1mm}
 \left[\begin{matrix}\cos^{2}{\left(\theta \right)} & \frac{i \sin{\left(2 \theta \right)}}{2} & 0 & 0 & 0 & 0 & - \sin^{2}{\left(\theta \right)} & \frac{i \sin{\left(2 \theta \right)}}{2}\\\frac{i \sin{\left(2 \theta \right)}}{2} & \cos^{2}{\left(\theta \right)} & \frac{i \sin{\left(2 \theta \right)}}{2} & - \sin^{2}{\left(\theta \right)} & 0 & 0 & 0 & 0\\- \sin^{2}{\left(\theta \right)} & \frac{i \sin{\left(2 \theta \right)}}{2} & \cos^{2}{\left(\theta \right)} & \frac{i \sin{\left(2 \theta \right)}}{2} & 0 & 0 & 0 & 0\\0 & 0 & \frac{i \sin{\left(2 \theta \right)}}{2} & \cos^{2}{\left(\theta \right)} & \frac{i \sin{\left(2 \theta \right)}}{2} & - \sin^{2}{\left(\theta \right)} & 0 & 0\\0 & 0 & - \sin^{2}{\left(\theta \right)} & \frac{i \sin{\left(2 \theta \right)}}{2} & \cos^{2}{\left(\theta \right)} & \frac{i \sin{\left(2 \theta \right)}}{2} & 0 & 0\\0 & 0 & 0 & 0 & \frac{i \sin{\left(2 \theta \right)}}{2} & \cos^{2}{\left(\theta \right)} & \frac{i \sin{\left(2 \theta \right)}}{2} & - \sin^{2}{\left(\theta \right)}\\0 & 0 & 0 & 0 & - \sin^{2}{\left(\theta \right)} & \frac{i \sin{\left(2 \theta \right)}}{2} & \cos^{2}{\left(\theta \right)} & \frac{i \sin{\left(2 \theta \right)}}{2}\\\frac{i \sin{\left(2 \theta \right)}}{2} & - \sin^{2}{\left(\theta \right)} & 0 & 0 & 0 & 0 & \frac{i \sin{\left(2 \theta \right)}}{2} & \cos^{2}{\left(\theta \right)}\end{matrix}\right]
\end{array}
\label{eq:U01}
\end{equation}
Matrix $\UU$ defines a cellular automaton. The pair of matrices $\UU_0$ and $\UU_1$ 
suggests a more fine-grained cellular evolution of the original Hamiltonian, 
by alternatingly applying $\UU_0$ and $\UU_1$ to the quantum state $\Psi(x)$.
An even-numbered cell interacts alternatingly with its right-hand side and left-hand side neighbors.
For odd-numbered cells the order of interaction is opposite.
The cell-update rules of $\UU$ have a bounded neighborhood,
including both immediate neighbors and one of the neighbors at distance two. 
Matrix $\UU$ has nonzero elements
in the top-right and bottom-left corners, consistent with the periodic boundary conditions.
Subsection \ref{subsec:X} presents a bounded, non-periodic, SCA.

Cellular automaton $\UU_1 \UU_0$ is a so-called {\em partitioning cellular automaton}, 
(a.k.a. block cellular automaton) 
pioneered by Margolus and Toffoli (\cite{1987-Toffoli}, pp 119-120).
The $N$ cells are partitioned in two ways, each with a matching {\em block rule}
($\UU_0$ and $\UU_1$), applied alternatingly.

The $ \mathcal{O}(\theta^2)$ term (\ref{eq:U_homogeneous})
is the so-called spectral-norm (approximation) error,
which can be made arbitrarily small by choosing a small time step $\tau$ (a small $\theta$).
This approximation can be improved \cite{1999-Sornborger} by
\begin{equation} 
\begin{array}{lll}
		 \expbig{-i \theta \Hhat}
	~=~ 	  \expbig{-i \frac\theta2 \Hhat_0} 
		~\expbig{-i \theta \Hhat_1} 
		~\expbig{-i \frac\theta2  \Hhat_0}  + \mathcal{O}(\theta^3)  ~.
\end{array}
\label{eq:U0U1U0}
\end{equation} 
The SCA experiments in later sections are based on the alternate application of 
$\UU_1 \UU_0 $ and $\UU_0 \UU_1$, according to this improved approximation.
Section \ref{sec:1D-experiments} proposes alternative measures for the approximation error,
in terms of (phase and group) velocities and energy levels.


\subsection{A 3-dimensional SCA for a free particle}
\label{subsec:3D}

A 3D SCA operates on wave function $\Psi $ 
represented by a 3D matrix of $N = N_X N_Y N_Z$ cells, with $N_X, N_Y, N_Z$ even.
This SCA can be represented by 2D matrix $\UU_{XYZ}$, such that
\begin{equation} 
     	\vecc \big(\Psi(t\!+\!\tau)\big)  =  \UU_{XYZ} \;\vecc \big(\Psi(t) \big) ~,
\label{eq:3D-goal}
\end{equation} 
where $\vecc(\MM)$ denotes the standard vectorization of matrix $\MM$.
(In two dimensions, $\vecc(\MM)$ is
obtained by stacking the columns of the  $\MM$ on top of one another.)
First, matrix $\UU_{XYZ}$ is derived from the Hamiltonian of a free particle,
 expressed in terms of 1D SCAs of (\ref{eq:U_homogeneous}).
In Subsection \ref{subsec:3D-potentials} an SCA 
for the complete Schr{\" o}dinger Hamiltonian (\ref{eq:schroedinger}) will be derived.

A 2D version of discrete Hamiltonian (\ref{eq:schroedinger-discrete}) is given by
\begin{equation}
\begin{array}{lll}
	\mathbf{H} \Psi(x,t) ~=~ - \delta_m \big(	&\Psi(x\!-\!1,y,t) - 2 \Psi(x,y,t) + \Psi(x\!+\!1,y,t)   \\
									&\Psi(x,y\!-\!1,t) - 2 \Psi(x,y,t) + \Psi(x,y\!+\!1,t) ~\big)  ~.
\end{array}
\label{eq:2D-discrete}
\end{equation} 
In matrix form, the Hamiltonian can be split as in $\HH_{XY}$,
where each of the two terms corresponds to a line of  (\ref{eq:2D-discrete}):
\begin{equation} 
\begin{array}{lll}
     	\HH_{XY}	&=&	\II_{X} \otimes \HH_{Y} ~+~  \KK_{XY} (\II_{Y} \otimes \HH_{X}) \KK_{XY}^\top 	\\[1mm]
	\HH_{X}  	&=&	\II_{X2} \otimes \BB  ~+~  \SB^{-1} (\II_{X2} \otimes \BB)  \;\SB		\\[1mm]
	\HH_{Y}  	&=&	\II_{Y2} \otimes \BB  ~+~  \SB^{-1} (\II_{Y2} \otimes \BB)  \;\SB ~.
\end{array}
\label{eq:2D-H}
\end{equation} 
Here identity matrix $\II_X$ is of size $N_X$ and $\II_{X2}$ of size $N_X/2$. 
Matrix $\KK_{XY}$ is a shorthand for the so-called {\em commutation matrix} $\KK^{(N_X,N_Y)}$
and is defined by  $\KK^{(m,n)} \vecc(\mathbf{M}) = \vecc(\mathbf{M}^\top)$ 
for $m \!\times\! n$ matrix $\mathbf{M}$.
Note that with  $\KK_{YX} \!=\! (\KK_{XY})^\top \!=\! (\KK_{XY})^{-1}$.
The two terms of $\HH_{XY}$ commute, that is, 
\begin{equation} 
     	\big[  \II_{X} \otimes \HH_{Y} ,  ~ \KK_{XY} (\II_{Y} \otimes \HH_{X}) \KK_{YX}^\top \big] ~=0 ~.
\label{eq:2D-commute}
\end{equation} 
Hence, since matrices $\KK$ are orthogonal,
\begin{equation} 
\begin{array}{lll}
	\exp\left(-i \frac\tau\hbar \HH_{XY}  \right)&=&
	\exp\left(-i \frac\tau\hbar \left( \II_{X} \otimes \HH_{Y}\right)\right) \cdot
	\exp\left(-i \frac\tau\hbar \left( \KK_{XY} (\II_{Y} \otimes \HH_{X} ) \KK_{XY}^\top \right) \right) 	\\[2mm]
	&=&
	\left( \II_{X} \otimes \exp\left(-i \frac\tau\hbar  \HH_{Y}\right)\right) \cdot
	\left( \KK_{XY} (\II_{Y} \otimes \exp\left(-i \frac\tau\hbar \HH_{X} ) \KK_{XY}^\top \right) \right) 	\\[2mm]
	&=& (\mathbf{I}_X \otimes \UU_Y) \cdot
			 \KK_{XY} \;(\II_Y \otimes \UU_X)\;  \KK_{XY} ^\top 	~+ \mathcal{O}(\theta^2)		\\[1mm]
	&=& \UU_X \otimes \UU_Y 							~+ \mathcal{O}(\theta^2)		
\end{array}
\label{eq:2D_U}
\end{equation} 
where 2D SCA  $~\UU_{XY} = \UU_X \otimes \UU_Y$ is constructed from 1D SCA (\ref{eq:U_homogeneous}):
\begin{equation} 
\begin{array}{lllll}
 	\UU_X = \UU_{X,1} \UU_{X,0} 
	&~~~~& \UU_{X,0} = \II_{X2} \otimes \CB
	&~~~~& \UU_{X,1} =\SB^{-1} (\II_{X2} \otimes \CB)  \;\SB	\\[1mm]
 	\UU_Y = \UU_{Y,1} \UU_{Y,0} 
	&& \UU_{Y,0} = \II_{Y2} \otimes \CB
	&& \UU_{Y,1} =\SB^{-1} (\II_{Y2} \otimes \CB)  \;\SB	~.
\end{array}
\label{eq:2D_U_U}
\end{equation} 
Alternatively, 
\begin{equation} 
	\UU_{XY} = \big(\UU_{X,1} \otimes \UU_{Y,1} \big) \cdot  \big(\UU_{X,0} \otimes \UU_{Y,0} \big) 	
\label{eq:2D-margolus}
\end{equation} 
which describes an SCA-evolution using the 2$\times$2 blocks of the Margolus neighborhood \cite{1987-Toffoli}.

For the 3D case, Hamiltonian $\HH_{XYZ}$  can be written as 
a sum of three mutually commuting terms, similar to (\ref{eq:2D-H}).
The resulting 3D SCA, using $\KK_{XYZ} = \KK^{(N_X,N_Y Nz)}$, is
\begin{equation} 
\begin{array}{lll}
	&& \UU_{XYZ} 													\\[1mm]
        &=& (\UU_X \otimes \UU_Y \otimes \UU_Z) 							\\[1mm]
	&=& 	\big(\II_X \otimes \II_Y \otimes \UU_Z \big) \cdot 
        		\big(\II_X \otimes \UU_Y \otimes \II_Z \big) \cdot    
		\big(\UU_X \otimes \II_Y \otimes \II_Z \big) 						\\[1mm]
	&=& \big(\II_X \otimes \II_Y \otimes \UU_Z \big) \cdot 
	\big(\II_X \otimes \KK_{YZ} \; (\II_Z \otimes \UU_Y) \; \KK_{YZ}^\top \big) \cdot 
        \big(\KK_{XYZ}  \; (\II_Y \otimes \II_Z \otimes \UU_X )\; \KK_{XYZ} ^\top \big) ~.
\end{array}
\label{eq:3D_U}
\end{equation}

\subsection{A 3-dimensional SCA, incl. $\phi(\xvec)$ and  $\AAA(\xvec)$}
\label{subsec:3D-potentials}

The complete Schr{\" o}dinger Hamiltonian (\ref{eq:schroedinger}) can be rewritten according to
\begin{equation} 
\begin{array}{lll}
&	\left(\frac{1}{2m} 	\left(\frac{\hbar}{i} \nabla - q \AAA(\xvec) \right) \cdot
				\left(\frac{\hbar}{i} \nabla - q \AAA(\xvec) \right) + q \phi(\xvec)\right) \Psi  			\\[2mm]
=&	\left(- \frac{\hbar^2}{2m} \nabla^2
	+ \left(q \phi(\xvec) + \frac{q^2}{2m} \AAA(\xvec)^2 \right)
	+ i\frac{q\hbar}{2m}\big(\nabla \!\cdot\! \AAA(\xvec) + \AAA(\xvec) \!\cdot\! \nabla \big)\right) \Psi 	\\[2mm]
=&	- \frac{\hbar^2}{2m} \nabla^2 \Psi 
	+ \left(q \phi(\xvec) + \frac{q^2}{2m} \sqabs{\AAA(\xvec)} \right) \Psi 
	+ i\frac{q\hbar}{2m} \big(\Psi  \nabla \!\cdot\! \AAA(\xvec) + 2 \AAA(\xvec) \!\cdot\! \nabla \Psi \big)  	\\[2mm]
=&	- \frac{\hbar^2}{2m} \nabla^2 \Psi 
	+ \left(q \phi(\xvec) + \frac{q^2}{2m} \sqabs{\AAA(\xvec)} \right) \Psi 
	+ i\frac{q\hbar}{2m} \big(2 \AAA(\xvec) \!\cdot\! \nabla \Psi \big)   ~.
\end{array} 
\label{eq:schroedinger_rewrite}
\end{equation} 
Here $\nabla \!\cdot\! \AAA(\xvec)$ is chosen to equal 0 (the Coulomb gauge), without loss of generality.
A discrete-space version of this Hamiltonian can be described by
\begin{equation} 
\begin{array}{lll}
\Psi(t+\tau) &=&~~~D_X  	(\xvec) \Psi(x\!-\!1, y, z,   t) 
                		~+~ 2C  	(\xvec) \Psi(\xvec,           t) 
                		~+~ D_X^*	(\xvec) \Psi(x\!+\!1, y, z,   t)  \\
             	&&  	  +~ D_Y   (\xvec) \Psi(x, y\!-\!1, z,   t) 
                		~+~ 2C 	(\xvec) \Psi(\xvec,           t) 
                		~+~ D_Y^*	(\xvec) \Psi(x, y\!+\!1, z,   t)  \\
             	&&  	  +~ D_Z 	(\xvec) \Psi(x,   y, z\!-\!1, t) 
                		~+~ 2C 	(\xvec) \Psi(\xvec,           t) 
                		~+~ D_Z^*	(\xvec) \Psi(x,  y, z\!+\!1, t) ~,
\end{array} 
\label{eq:2D-H-VA-discrete}
\end{equation} 
with, for $W \in \{X,Y,Z\}$
\begin{equation} 
\begin{array}{llll}
C(\xvec) 		&=& \delta_m + \frac13 \frac12 \big( q \phi(\xvec) 
						+ \frac{\delta_q^2}{\delta_m} \sqabs{\AAA(\xvec)}\big)  \\[2mm]
D_W(\xvec)	&=& -\delta_m - 2i \delta_q A_W(\xvec)    ~,     
\end{array}
\label{eq:2D-H-VA-discrete-1}
\end{equation} 
where $\delta_m$ (as before) and $\delta_q$ are given by
\begin{equation} 
	\delta_m = \frac1{2m}\frac{\hbar^2}{a^2}   ~,    \hspace{2cm}
	\delta_q = \dfrac{q}{2m}\dfrac{\hbar}{a} ~,   \hspace{2cm}
	\frac{\delta_q^2}{\delta_m} = \frac{q^2}{2m} ~.
\label{eq:2D-H-VA-discrete-2}
\end{equation} 
The factors $\frac13$ and $\frac12$ correspond to the 3D split and the even/odd splits of the Hamiltonian.
In matrix form, it can be expressed in terms of 1D split Hamiltonians, such as
\begin{equation} 
\begin{array}{llll}
	\HH_X(\xvec) 
	&=& \bigoplus_{x=0}^{N_X/2} \BB_{X}(2x, y,z) ~+~
		\mathbf{S}^{-1} \left(\bigoplus_{x=0}^{N_X/2} \BB_{X}(2x\!+\!1 \!\!\!\!\mod N_X, y,z)\right) \mathbf{S} \\[2mm]
	\BB_W(\xvec)  
	&=& \left[\begin{matrix}
		C  		(\xvec) & D_W	(\xvec)  \\
		D_W^*	(\xvec) & C	(\xvec)
    		\end{matrix}\right] ~,
\end{array}
\label{eq:1D-H-VA}
\end{equation} 
where $\oplus$ denotes the {\em direct matrix sum}, not to be confused with the Kronecker sum.
The expressions for $\HH_Y(\xvec)$ and $\HH_Z(\xvec)$ are similar.
For 3D Hamiltonian $\HH_{XYZ}$
each of the three terms corresponds to a line of  (\ref{eq:2D-H-VA-discrete}):
\begin{equation} 
\begin{array}{llll}
	\HH_{XYZ} 
	&=& ~~~\left(\bigoplus_{x=0}^{N_X-1} \bigoplus_{y=0}^{N_Y-1} ~~~~~~~~~~~\HH_Z(\xvec) \right) \\[2mm]
       	&&+	\left(\bigoplus_{x=0}^{N_X-1} \KK_{YZ} 
        		\left(\bigoplus_{z=0}^{N_Z-1} ~~\; \HH_Y(\xvec) \right)  \KK_{YZ} ^\top \right) \\[2mm]
	&&+ \left(\KK_{XYZ}  \left(\bigoplus_{y=0}^{N_Y-1} 
		\bigoplus_{z=0}^{N_Z-1} ~\HH_X(\xvec) \right) \KK_{XYZ} ^\top \right) ~.
\end{array}
\label{eq:2D-H-VA}
\end{equation} 
Unlike Hamiltonian (\ref{eq:2D-H}) 
the matrix terms of (\ref{eq:2D-H-VA}) do {\em not} commute with one another.
Three-dimensional Schr{\" o}dinger cellular automaton $\UU_{XYZ}$ is given by
\begin{equation} 
\begin{array}{lll}
	\exp\left(-i \frac\tau\hbar \HH_{XYZ} \right)  
	&=&   \UU_{XYZ} ~+ \mathcal{O}(\tau^2)  \\[2mm]
	\UU_{XYZ} 
	&=&	\left(\bigoplus_{x=0}^{N_X-1} \bigoplus_{y=0}^{N_Y-1}  ~~~~~~~~~~
		~\;	\UU_{Z1}(\xvec) ~ \UU_{Z0}(\xvec)  \right)  \\[2mm]
	&&	\left(\bigoplus_{x=0}^{N_X-1} \KK_{YZ} \; \left(\bigoplus_{z=0}^{N_Z-1} 
		~~\;	\UU_{Y1}(\xvec) ~ \UU_{Y0}(\xvec)  \right) \; \KK_{YZ} ^\top \right)  \\[2mm]
	&& 	\left(\KK_{XYZ} \; \left(\bigoplus_{y=0}^{N_Y-1} \bigoplus_{z=0}^{N_Z-1} 
		~  	\UU_{X1}(\xvec) ~ \UU_{X0}(\xvec)  \right)\; \KK_{XYZ} ^\top \right)  ~,
\end{array}
\label{eq:3D_U-VA}
\end{equation} 
where $\UU_{X0}(\xvec)$ and $\UU_{X1}(\xvec)$ are defined by 
\begin{equation} 
\begin{array}{llll}
	\exp\left(-i \frac\tau\hbar \HH_{X}(\xvec) \right)  
					&=&	\UU_{X1}(\xvec) \cdot \UU_{X0}(\xvec) ~+ \mathcal{O}(\tau^2)	\\[2mm]
	\UU_{X0}(\xvec)	&=&	\bigoplus_{x=0}^{N_X/2}   \CB_X(2x, y, z) 					\\[2mm] 
   	\UU_{X1}(\xvec) 	&=& \SB^{-1} \!\left(	\bigoplus_{x=0}^{N_X/2}   
						\CB_{X}(2x\!+\!1 \!\!\!\!\mod N_X, y, z)  \right)\! \SB 			\\[2mm]
	\CB_W(\xvec)	 	&=&  \exp[\big(\!-\! i \frac\tau\hbar \;\BB_{W}(\xvec)\big)~,
\end{array}
\label{eq:1D-U-VA}
\end{equation} 
and likewise for $\UU_{Y0}$, $\UU_{Y1}$, $\UU_{Z0}$, and $\UU_{Z1}$.
As with (\ref{eq:U0U1U0}), the approximation error can be reduced to $\mathcal{O}(\tau^3)$
by suitably alternating the evolution order of the six $U_{Wi}$ involved.

\subsection{Reflective cells $X(\xvec)$}
\label{subsec:X}

How can a position-dependent potential-energy function $V(\xvec) = q\phi(\xvec)$ 
be used to introduce boundary conditions for $\Psi(\xvec)$?
In a one-dimesional SCA, $V(x)$ can be included in the Hamiltonian matrix as in
\begin{equation} 
	\Hhat' = \Hhat + \Vhat  \hspace{2cm} 
	\Vhat  = \diag\left(\frac1\delta_m V(x,y)\right) ~,
\label{eq:H_potential}
\end{equation} 
corresponding to the evolution matrix
\begin{equation} 
\begin{array}{llll}
	 \UU' = \exp{(-i \theta (\Hhat +\Vhat ))} 
	 &=& \exp{(-i \theta \Hhat )} \exp{(-i \theta \Vhat )}  
	  	&+ \mathcal{O}(\theta^2) \\[1mm]
	 &=& \exp{(-i \theta \Hhat )} \exp{\left(-i \theta (\Vhat + 2\frac{n}{\theta}\pi)\right)}  
	 	&+ \mathcal{O}(\theta^2)~.
\label{eq:HV}
\end{array}
\end{equation} 
Note that $\UU'$ is periodic in the potential, which limits the use of $V(x)$.
In particular, it can be shown that no real-valued $V(x)$ 
can cancel the off-diagonal nonzeros of matrix $\UU_1 \UU_0$ in (\ref{eq:U01}). 
Hence, $V(x)$ cannot be used to remove the periodic boundaries.

For a bounded SCA, one with reflective (non-periodic) boundaries, 
matrix $\mathbf{U}_0 = \mathbf{I}_M \otimes \mathbf{C} $ in (\ref{eq:U_homogeneous})
can be replaced by one where for the last block rotation angle $\theta \!=\! \pi$, as in
\begin{equation} 
 \mathbf{U}_{b1} = 
 \SB^{-1} \diag \big( \mathbf{C}, \;  \mathbf{C}, \;\ldots, \; \mathbf{C},  \;  \exp(-i \pi \BB) \big) \;\SB~.
\label{eq:reflective}
\end{equation} 
For $N\!=\!8$
\begin{equation} 
\begin{array}{ll}
\mathbf{U}_{b1}   = &\expbig{-i\theta} \times \\ \vspace{1mm}
& \left[\begin{matrix}- e^{i \theta} & 0 & 0 & 0 & 0 & 0 & 0 & 0\\0 & \cos{\left(\theta \right)} & i \sin{\left(\theta \right)} & 0 & 0 & 0 & 0 & 0\\0 & i \sin{\left(\theta \right)} & \cos{\left(\theta \right)} & 0 & 0 & 0 & 0 & 0\\0 & 0 & 0 & \cos{\left(\theta \right)} & i \sin{\left(\theta \right)} & 0 & 0 & 0\\0 & 0 & 0 & i \sin{\left(\theta \right)} & \cos{\left(\theta \right)} & 0 & 0 & 0\\0 & 0 & 0 & 0 & 0 & \cos{\left(\theta \right)} & i \sin{\left(\theta \right)} & 0\\0 & 0 & 0 & 0 & 0 & i \sin{\left(\theta \right)} & \cos{\left(\theta \right)} & 0\\0 & 0 & 0 & 0 & 0 & 0 & 0 & - e^{i \theta}\end{matrix}\right] ~.
\end{array}
\label{eq:U_inhomogeneous}
\end{equation} 
Crucially, $\UU_{b1}  [0,7] = \UU_{b1}  [7,0] = 0$, unlike in (\ref{eq:U1}).
Matrix $\UU_{b1} $ is {\em not} circulant, 
and the corresponding cellular automaton $\UU_{b1}  \UU_{b0}$ has no periodic boundaries, 
but a reflective (two-sided) boundary at position $N\!-\!1$.
Such {\em reflective cells} can be placed at any position $\xvec$, 
and can thus be used to model, for example,
an infinite potential well (Subsection \ref{subsec:infinite_well}),
or a double-slitted screen (Subsection \ref{subsec:double-slit}).

\subsection{A Schr{\" o}dinger quantum walk}
\label{subsec:qw}

A one-dimensional Schr{\" o}dinger QW for a free particle, comprising  $2M$ cells, 
and with periodic boundary conditions is presented next.
State vector $\qstate{\Psi}$ can be split into even and odd cells, as in
\begin{equation} 
   \qstate{\Psi} =\sum_{n=0}^{M-1}(\gamma_{2n} \qstate{2n} +\gamma_{2n+1} \qstate{2n+1} )~,
\label{eq:1D-qw-psi}
\end{equation} 
where $\gamma_i$ is a complex number. This state can be ``curled up" as in \cite{1993-Aharonov}
\begin{equation} 
    \qstate{\widetilde{\Psi}}
    = \sum_{n=0}^{M-1}(\alpha_{n} \qstate{\uparrow} +
             			      \beta_{n} \qstate{\downarrow})\otimes |n \rangle ~,
\hspace{15mm}
\alpha_n=\gamma_{2n} ~~\text{and}~~ \beta_n=\gamma_{2n+1} ~,
\label{eq:1D-qw-psi-2}
\end{equation} 
where two adjacent cells are combined into a single, non-normalized, qubit.
A corresponding two-step QW evolution operator can be described by matrix $\widetilde{\UU}$
\begin{equation} 
	\widetilde{\Psi}(t+1) ~=~ \widetilde{\UU} \; \widetilde{\Psi}(t) ~,
	\hspace{2cm}
	\widetilde{\UU} ~=~ \widetilde{\SB}^{-1}(\CB \otimes \IB_M) \; \widetilde{\SB} \; (\CB\otimes \IB_M) ~.
\label{eq:1D-qw-U}
\end{equation} 
Here $\CB$ is the so-called coin, which for the Schr{\" o}dinger QW is given by (\ref{eq:U_homogeneous}).
QW-shifts $\widetilde{\SB}$ and its inverse are given by
\begin{equation} 
  \begin{array}{llll}
    \widetilde{\SB} &=&\Sigma_{n=0}^{M-1} ~        
    	(\qstate{\downarrow} \qqstate{\uparrow} \otimes \qstate{(n\!-\!1)_M} \qqstate{n_M}	&+~
	 \qstate{\uparrow} \qqstate{\downarrow} \otimes \qstate{n_M} \qqstate{n_M} ) 		\\[2mm]
    \widetilde{\SB}^{-1} &=& \Sigma_{n=0}^{M-1} ~
    	(\qstate{\downarrow} \qqstate{\uparrow} \otimes \qstate{n_M} \qqstate{n_M} 		&+~
	 \qstate{\uparrow} \qqstate{\downarrow} \otimes \qstate{(n\!+\!1)_M} \qqstate{n_M} )	~,  
  \end{array}
\label{eq:1D-qw-shift}
\end{equation} 
where  $n_M$ is a shorthand for $n\bmod M$.
Shift operator $\widetilde{\SB}$ keeps the down-part of the state at its position and moves the up-part to the left. 
The inverse operator $\widetilde{\SB}^{-1}$ does the opposite. 

The QW in the form (\ref{eq:1D-qw-U}) can be found as staggered QW in \cite{2015-Portugal} 
and as alternating QW in \cite{Bru-2016}.
As shown Costa et al \cite{2018-Costa}, staggered QWs are a subset of PUQCAs.
Indeed, the QW above is structurally very similar to the evolution operator 
$\mathbf{U}_1 \mathbf{U}_0$ of (\ref{eq:U_homogeneous}),
\begin{equation} 
	\UU ~=~ \UU_1 \UU_0 ~=~  \SB^{-1} (\IB_M \otimes \CB) \;\SB \; (\IB_M \otimes \CB) ~.
\label{eq:U1U0_CA}
\end{equation} 
QWs typically use different coins $\CB$, including Hadamard, Fourier, and Grover coins.
Alternatively, the SCA based on these coins also specify unitary cellular automata.

The SCA and the staggered QW are like two sides of the same medal. 
A CA appears to be more natural as a solution of the Schr{\" o}dinger equation, 
while a QW is a more natural way to model search. 
It is comparable with the classical Markov processes, and the backward and forward Kolmogorov equations: 
the forward equations, also called Fokker-Planck equations, are used to describe the evolution of the probabilty distribution, 
which is perfectly suited for a CA, 
while the backward equations are used to compute the probability to reach a certain state.

\newpage

\section{Experiments with 1D Schr{\" o}dinger cellular automata}
\label{sec:1D-experiments}

In this section the results of the evolution of 1D SCA
are compared with known analytical solutions for various textbook Hamiltonians. 

\subsection{A plane wave in a periodic SCA }
\label{subsec:plane_wave}

In a first experiment a 1D SCA is initialized with a plane wave.
Cell $x$ has position $xa$ for integer cell index $x$ and cell size $a$.
A plane wave at time $t\!=\!0$ as a function of  $x$, 
with spatial offset $x_0$ and wavenumber $k$, can be described by $W_{\mathit plane}$
\begin{equation} 
  W_{\mathit plane} (x,x_0,k) = e^{i k (x-x_0)} ~.
  \label{eq:W_plane}
\end{equation} 
Note that also wavenumber $k$ is scaled by cell size $a$ such that $k = a 2\pi/\lambda$, 
and as a result is dimensionless. 
An initial state of a cellular automata can be obtained by sampling $W_{\mathit plane}$ at each individual cell index. 
A given cell size $a$ implies a shortest possible wavelength, viz. $\lambda \!=\! 2a$, 
which corresponds to $k\!=\!\pi$. 
Sampling of $W_{\mathit plane}$ at discrete positions leads to a form of spatial aliasing, since, for integer $n$
\begin{equation} 
  W_{\mathit plane}(x,x_0,k+ 2n\pi) = W_{\mathit plane}(x,x_0,k) ~.
  \label{eq:W_plane_alias}
\end{equation} 
This aliasing has important implications for phase velocity, dispersion,  group velocity, and energy levels, as is explored below.

\begin{figure}[!ht]
  \begin{center}
    \includegraphics[width=0.999\textwidth]{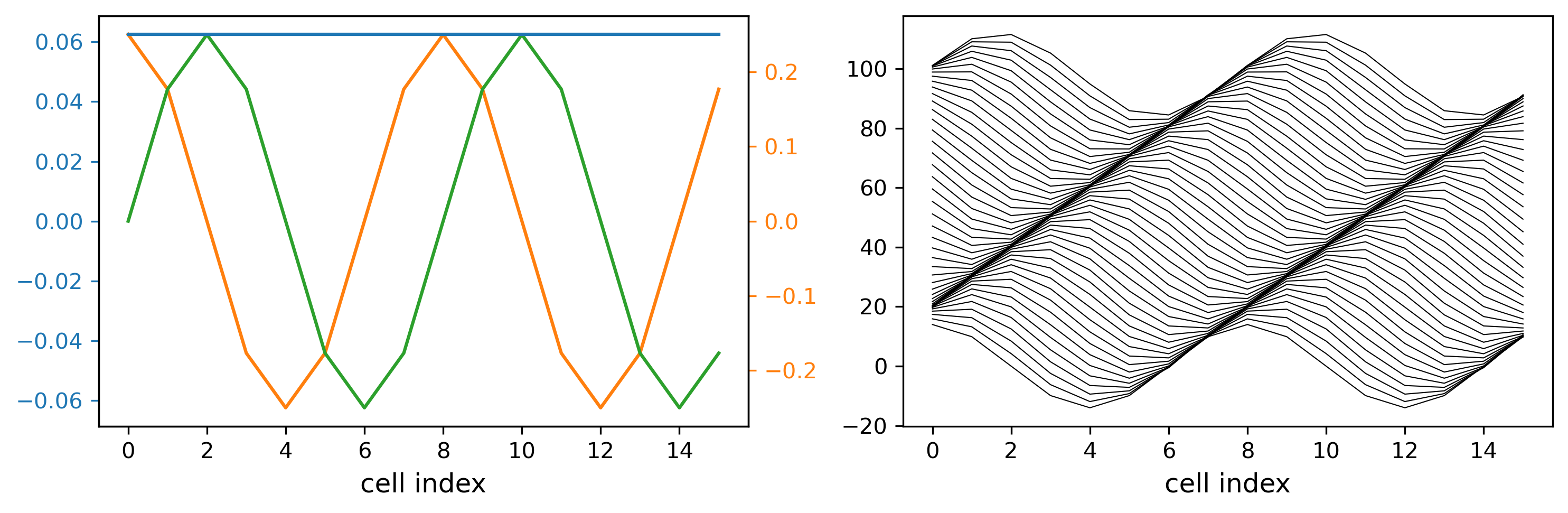}
  \end{center}
  \vspace{-5mm}
\caption{
  (left)  $\Psi(x,t\!=\!0)$ is plane wave with $\RE(\Psi)$ in orange, $\IM(\Psi)$ in green, and $\sqabs{\Psi}$ in blue;
  (right) the plane wave moves rightwards over 100 iterations, two at a time.
}
\label{fig:periodic_wave_time}
\end{figure}

The first cellular automaton comprises 16 cells with $\theta\!=\!\frac{\pi}{24}$.
It is initialized with $W_{\mathit plane}$ with $k$ chosen such that precisely two wavelengths fit in the 16 cells.
Figure \ref{fig:periodic_wave_time} shows this initial state and its evolution over 100 cycles.
The integrated probability $P(t)$ remains very close to $1$. 
After as many as 100k cycles $\mid\! P\!-\! 1 \!\mid \;\approx 10^{-11}$, without any form of normalization.
The plane wave moves rightwards with a phase velocity of about 10 cells per 100 cycles, $\approx 0.1$ $[a/\tau]$.
Below the the phase velocity is calculated 
as {\em half} the displacement of a plane wave over {\em two} evolution steps,
based on the alternate application of $\mathbf{U}_{1}\mathbf{U}_{0}$ and $\mathbf{U}_{0}\mathbf{U}_{1}$.
The application of two evolution steps to $\Psi(x, t_0) = e^{i k x}$ results in
\begin{equation} 
\begin{array}{llll}
\Psi(x, t_0\!+\!1) &=& (\mathbf{U}_1 \mathbf{U}_0) \cdot \Psi(x, t_0)  \\[1mm]
  &=&  e^{i k x} e^{-2i \theta} \big(
        - \!\sin^2(\theta)  e^{-2i k} 
        + \frac12 i\sin  (2\theta) e^{- i k} 
        + \cos^2(\theta)                           
        + \frac12 i\sin  (2\theta) e^{  i k} \big)\\[1mm]    
  &=& e^{i k (x-v_{10}) } ~,  \\
\mathrm{where}~~ v_{10} &=& \frac1{k}  \Big(2\theta - \ln \big(
          \cos^2(\theta)                           
        - \sin^2(\theta)  e^{-2i k} 
        +i\sin(2\theta) \cos(k) 
        \big) \Big) ~. \\[2mm]  
\Psi(x, t_0\!+\!2) &=& (\mathbf{U}_0 \mathbf{U}_1) \cdot \Psi(x, t_0\!+\!1)  
                           ~=~ (\mathbf{U}_0 \mathbf{U}_1) \cdot e^{i k (x-v_{10} )}\\[1mm]  
  &=& e^{i k (x-v_{10} ) } e^{-2i \theta} \big(
        - \!\sin^2(\theta)  e^{2i k} 
        + \frac12 i\sin  (2\theta) e^{- i k} 
        + \cos^2(\theta)                           
        + \frac12 i\sin  (2\theta) e^{  i k} \big)\\[1mm]    
  &=& e^{i k\left(x-(v_{10}+v_{01}) \right) } ~,   \\[1mm]
\mathrm{where}~~ v_{01} &=& \frac1{k}  \Big(2\theta - \ln \big(
          \cos^2(\theta)                           
        - \sin^2(\theta)  e^{2i k} 
        +i\sin(2\theta) \cos(k) 
        \big) \Big)~. 
\end{array}             
  \label{eq:v_p0}
\end{equation} 
The Taylor expansion of both $\ln$ functions, assisted by Python/Sympy, results in
\begin{equation} 
\begin{array}{lll}
   v_{10}
   &=& ~~\dfrac{4 \theta}{k} \sin^2{\left(\frac{k}2 \right)}
       + \dfrac{\theta^{2}}{k} \sin{\left(k \right)}
       + \dfrac{4\theta^{3}}{3k} \Big( \cos{\left(k \right)} \sin^2\left(k \right) -
                                    3i \sin{\left(k \right)} \cos^2\left(k \right)
                                 \Big) \\[2mm]
   &&  - \dfrac{\theta^{4}}{6k} \big(8\sin \left(2k \right) 
                                     +3\sin \left(4k \right) \big)
       +  \mathcal{O} \left(\theta^{5}\right) \\[4mm]
   v_{01}
   &=& ~~\dfrac{4 \theta}{k} \sin^2{\left(\frac{k}2 \right)}
       - \dfrac{\theta^{2}}{k} \sin{\left(k \right)}
       + \dfrac{4\theta^{3}}{3k} \Big( \cos{\left(k \right)} \sin^2\left(k \right) +
                                    3i \sin{\left(k \right)} \cos^2\left(k \right)
                                 \Big) \\[2mm]
   &&  + \dfrac{\theta^{4}}{6k} \big(8\sin \left(2k \right) 
                                     +3\sin \left(4k \right) \big)
       +  \mathcal{O}\left(\theta^{5}\right) ~.
\end{array}
  \label{eq:v_p1}
\end{equation} 
Note that both $v_{10}$ and $v_{01}$ are complex and that
the imaginary terms as well as the (real)  $\theta^{2}$ and $\theta^{4}$ terms  in $v_{10}$ and $v_{01}$
have opposite signs.
Finally, phase velocity $v_p(\theta, k)$ is given by
\begin{equation} 
\begin{array}{lll}
   v_p(\theta, k)
   &=&   \frac12 \big(v_{10} + v_{01}\big) \\[2mm]
   &=&     \dfrac{4 \theta}{k} \sin^2{\left(\frac{k}2 \right)}
	   + \dfrac{4 \theta^{3}}{3k} \cos{\left(k \right)}\sin^2{\left(k \right)}
    	   + \mathcal{O} \left(\theta^{5}\right) ~, 
\end{array}
  \label{eq:v_p2}
\end{equation} 
where the imaginary, $\theta^{2}$, and $\theta^{4}$ contributions cancel each other.
This cancelation is a direct consequence of the alternating application of
$\mathbf{U}_1 \mathbf{U}_0 $ and $\mathbf{U}_0 \mathbf{U}_1$
and is consistent with the spectral-norm approximation error in  (\ref{eq:U0U1U0}).
For the experiment of Figure \ref{fig:periodic_wave_time} this results in $v_p(\frac\pi{24}, \frac\pi{4} ) = 0.098$,
 in agreement with the observed phase velocity.
 
\begin{figure}[!ht]
  \begin{center}
    \includegraphics[width=0.999\textwidth]{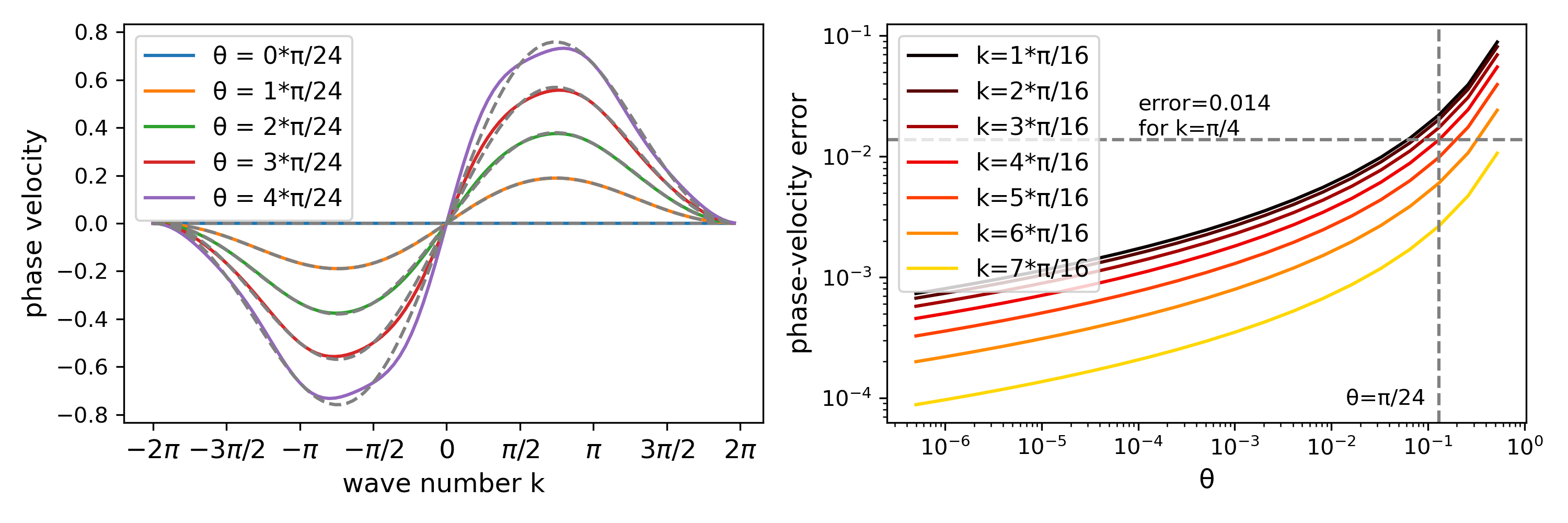}
  \end{center}
  \vspace{-5mm}
\caption{
(left) Phase velocity $v_p$  versus wave number k, $k \in [-2\pi, 2\pi)$, in units of $\frac{a}{\tau}$
for various $\theta$.
The coloured curves  show  the actual phase velocity $v_p(\theta, k)$ according to (\ref{eq:v_p2}).
The gray dashed curves show the base phase velocity ${\bf v}_p(\theta, k)$ according to (\ref{eq:v_p-base}).
(right) The relative error in phase velocity versus $\theta$ according to (\ref{eq:v_p-error}).
}
\label{fig:phase_velocity}
\end{figure}

The Lie-Trotter product formula applied to (\ref{eq:U_homogeneous}) gives
\begin{equation} 
	  \expbig{-i \theta (\Hhat_1 + \Hhat_0)}
  = \lim_{n\rightarrow \infty} \bigg( \expBig{-i \frac\theta{n} \Hhat_1} 
  						  \expBig{-i \frac\theta{n}  \Hhat_0}   \bigg)^n ~.
  \label{eq:lie-trotter}
\end{equation} 
The approximation error in $v_p(\theta, k)$ can be reduced by considering
$n$ smaller displacements over time $\tau/n$.
This suggests the following definition of the {\em base} phase velocity ${\bf v}_p(\theta, k)$.
\begin{equation} 
{\bf v}_p(\theta, k) ~=~ \lim_{n \rightarrow \infty} n v_p \Big(\frac\theta{n}, k\Big) 
		             ~=~ \dfrac{4 \theta}{k} \sin^2{\left(\frac{k}2 \right)}  ~.
  \label{eq:v_p-base}
\end{equation} 
This base phase velocity corresponds to the phase velocity of the cellular automaton in the limit 
$\tau \!\rightarrow \! 0$.
Hence, the $\mathcal{O}\left(\theta^{3}\right)$ term in (\ref{eq:v_p2})
{\em is} the absolute error in phase velocity $v_p$ as introduced by splitting the Hamiltonian
according to (\ref{eq:U0U1U0}).\footnote{
The relation between this error measure and the eigenvalue error \cite{2024-Morales}
is a topic for further study.}
The {\em relative} error in phase velocity $v_p$ then becomes
\begin{equation} 
\frac{v_p(\theta, k) - {\bf v}_p(\theta, k)}{{\bf v}_p(\theta, k)} 
~=~ \theta^2 \frac{4 \cos{\left(k \right)}\sin^2{\left(k \right)} }       
                 {3 \sin^2{\left(\frac{k}2 \right)}} + \mathcal{O}\left(\theta^{4}\right) ~.
  \label{eq:v_p-error}
\end{equation} 
Figure \ref{fig:phase_velocity} depicts the  phase and base velocities versus $k$,
as well as this relative error.

\subsection{A Gaussian wave packet in a periodic SCA}
\label{subsec:wave-packet }

In the next experiment the same cellular automaton
is initialized with a Gaussian wave packet of the form $W_{\mathit packet}$,
using a normalization constant $A$.
\begin{equation} 
	W_{\mathit packet}(x,x_0,\sigma,k) = 
	A\; \exp \left( -\frac12 \big(\frac{x-x_0}{\sigma} \big)^2 \right) 
	\exp \big(i k (x-x_0) \big) ~.
  \label{eq:W_packet}
\end{equation} 

\begin{figure}[!h]
     \centering
     \begin{subfigure}[b]{\textwidth}
         \centering
    \includegraphics[width=0.999\textwidth]{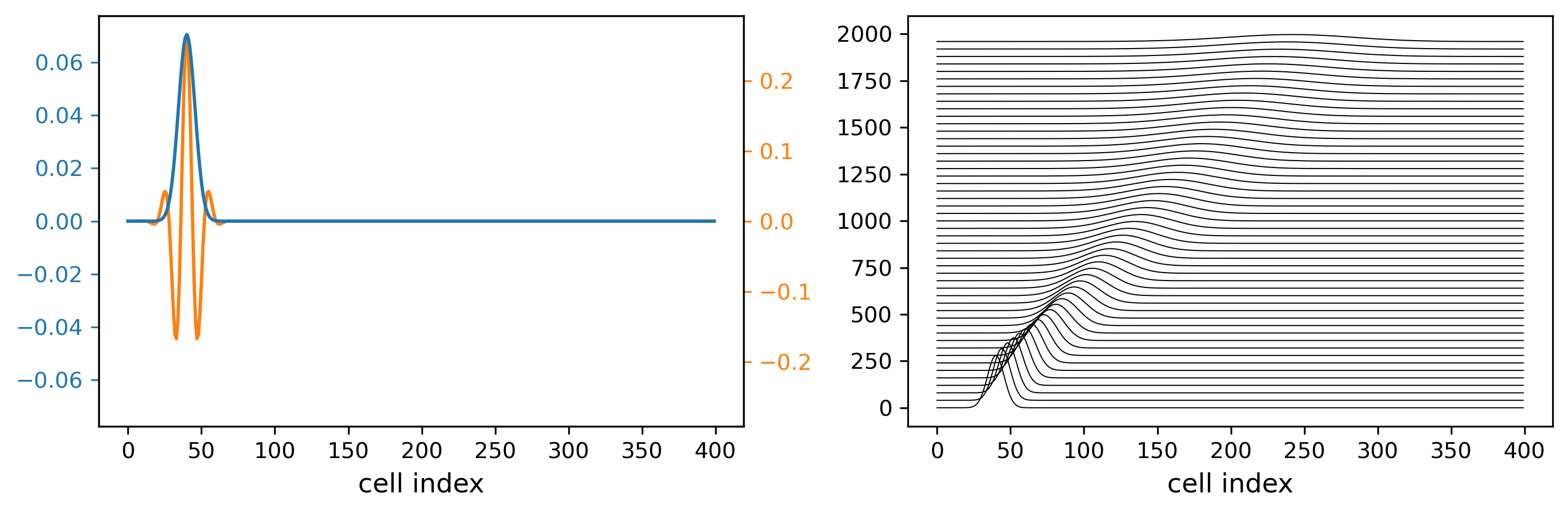}
         \caption{Wavenumber $k =\pi/8$: the wave packet is ``sparse'' and quickly disperses.}
         \label{fig:wavepackets-sparse}
     \end{subfigure}
     \vspace{3mm}
     
     \begin{subfigure}[b]{\textwidth}
         \centering
    \includegraphics[width=0.999\textwidth]{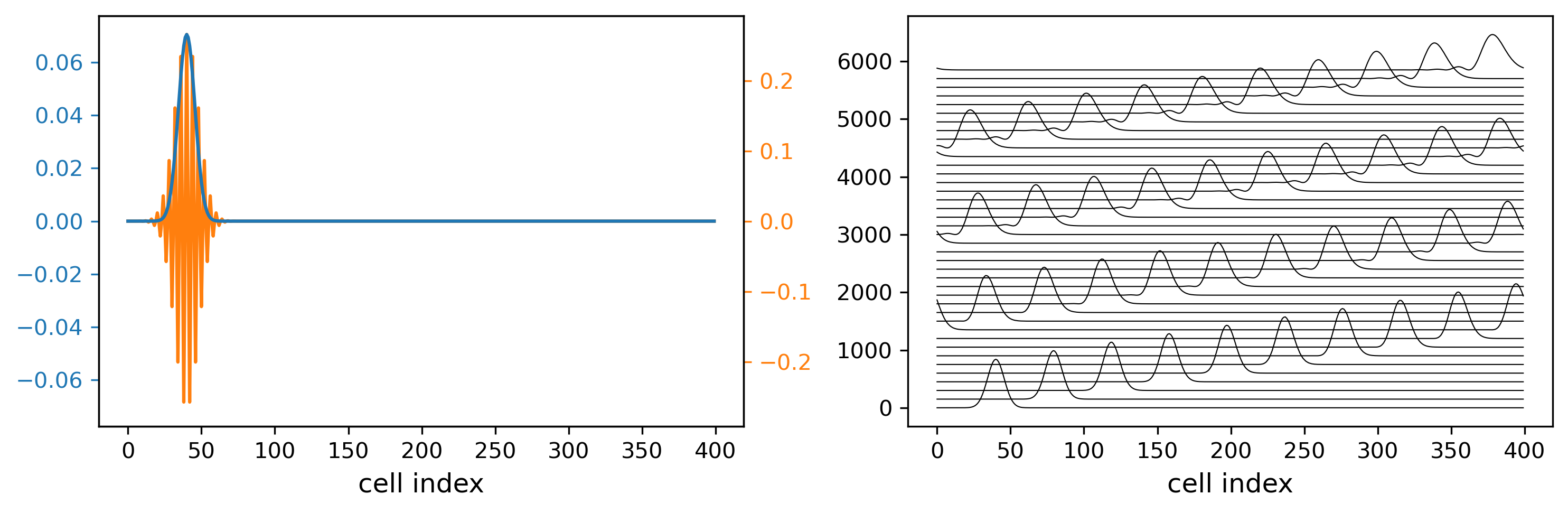}
         \caption{Wavenumber $k =\pi/2$: the wave packet is ``dense" and disperses much slower.}
         \label{fig:wavepackets-dense}
     \end{subfigure}
\caption{A Gaussian wave packet moves rightwards in a periodic SCA of 400 cells.}
\end{figure}

Figure \ref{fig:wavepackets-sparse} shows a wave packet $W_{\mathit packet}$, 
centered around position $x_0\!=\!40$ with $\sigma\!=\!8$ and $k \!=\! \frac{\pi}{8}$.
Note that the spatial extent of the wave packet appears to be less than three wavelengths.
The wave packet moves rightwards rather slowly, fewer than 200 cells in 2000 cycles.
Furthermore, it rapidly disperses. 
Figure \ref{fig:wavepackets-dense} shows the same automaton 
initialized with a different wave packet with $\sigma\!=\!8$ and $k \!=\! \frac{\pi}{2}$.
Roughly ten wavelengths fit in the Gaussian envelope.
It takes about 1600 cycles for a packet to traverse the width of the cellular automaton, 
and to reappear at the left hand side. 
The measured group velocity is $\approx$ 0.26 cells per cycle.
After 8000 cycles the dispersion of the wave packet is visible
as a minor pulse trailing the main wave packet.

A wave packet is composed of component sinusoidal waves of different wave numbers, 
each propagating with its own wavenumber-dependent phase velocity.
The wave packet in Figure \ref{fig:wavepackets-sparse}  has a center $k \!=\! \frac{\pi}{8}$.
Component waves with a lower $k$ move slower and those with a higher $k$ faster.
The dispersion of the second wave packet is substantially less,
because of the smaller bandwidth $\Delta k/k$ of the wave packet.
The component wave of the center frequency ($k\!=\! \frac\pi2$) moves fastest;
all the other component waves move slower.
(Recall that with $k\!=\! \frac\pi2$ the phase velocity of the center frequency of the wave packet is zero.)

It is common to derive the phase velocity from the so-called {\em dispersion relation} $\omega(k)$. 
Conversely, from (\ref{eq:v_p2}) it follows that
\begin{equation} 
  \omega(\theta,k) ~=~ k v_p(\theta,k)
      		 ~=~    4 \theta \sin^2{\left(\frac{k}2 \right)}
       			+ \frac43 \theta^{3} \cos{\left(k \right)}\sin^2{\left(k \right)}
    			+ \mathcal{O} \left(\theta^{5}\right) ~.
  \label{eq:dispersion_relation}
\end{equation} 
The phase velocity $v_p$ and group velocity $v_g$ are related by
\begin{equation} 
  v_p = \frac\omega{k} \hspace{2cm} \mathrm{and} \hspace{2cm}
  v_g = \frac{d\omega}{dk} ~. 
    \label{eq:group_velocity_definition}
\end{equation} 
Then from (\ref{eq:v_p2}) it follows that
\begin{equation} 
   	   v_g(\theta, k)
   ~=~  \frac{d (k v_p)}{dk}  
   ~=~ 	2 \theta \sin{\left(k \right)}  
        	     + \frac43 \theta^{3} \Big(2 \sin{\left(k \right)} - 3 \sin^{2}{\left(k \right)} \Big)
     	     +  \mathcal{O} \left(\theta^{5}\right)   ~.       
\label{eq:group_velocity}
\end{equation} 
As with the phase velocity, the $\mathcal{O} \left(\theta^{3}\right)$ contribution 
is the (absolute) approximation error.

\begin{figure}[!ht]
  \begin{center}
    \includegraphics[width=0.999\textwidth]{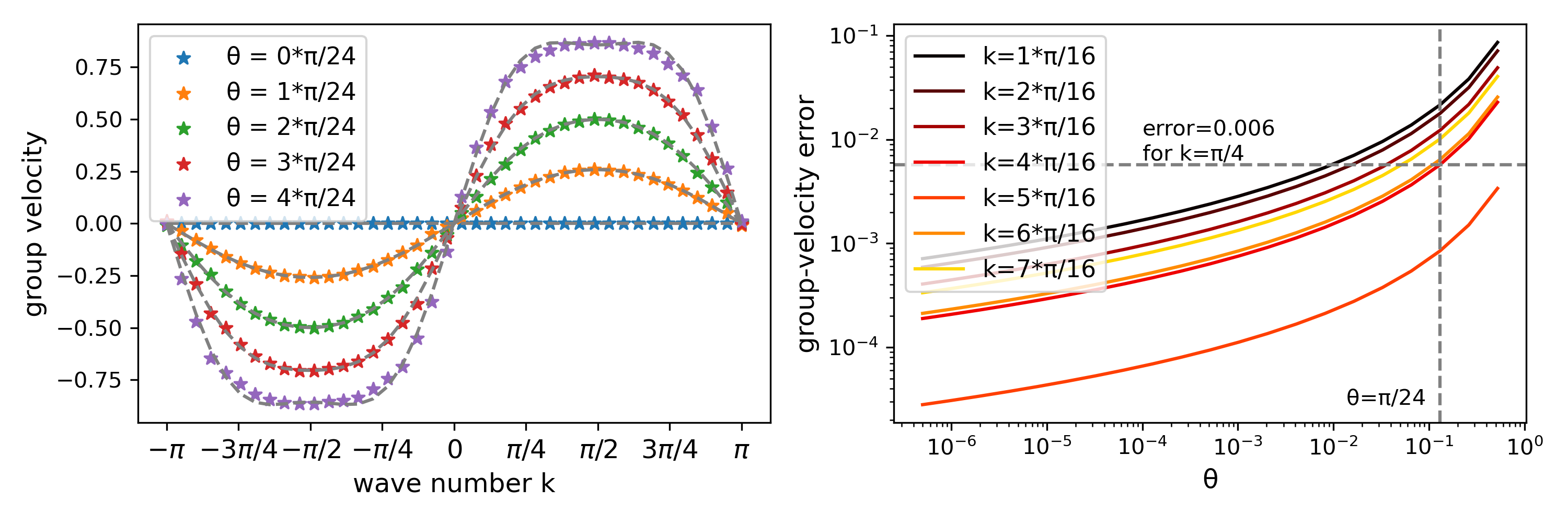}
  \end{center}
  \vspace{-5mm}
\caption{
(left) Group velocity $v_g$ versus wave number $k$  for various $\theta$ in units of $\frac{a}{\tau}$.
The colored $\star$ markers denote {\em measured} group velocities from SCA evolutions.
The dashed gray lines are based on (\ref{eq:group_velocity}).
(right) The relative error in group velocity versus $\theta$.
}

\label{fig:group_velocity}
\end{figure}

Figure \ref{fig:group_velocity} shows the measured group velocity $v_g$ versus $(\theta, k)$. 
These measurements are based on the displacement of the peak of the Gaussian envelope of the wave packet 
during a time interval of 200 cycles.
The group velocity is periodic in $k$, approximately sinusoidal.
It is well known that imposing periodic boundary conditions in space leads to discrete momenta. 
It is less wel known is that discretization of space leads to periodic momenta. 
The momentum interval $[-\pi , \pi)$, used to depict $v_g(k)$, is known as the Brillouin zone.

\subsection{A Gaussian wave packet in a bounded SCA }
\label{subsec:well_packets}

\begin{figure}[!h]
     \centering
     \begin{subfigure}[b]{\textwidth}
         \centering
    \includegraphics[width=0.999\textwidth]{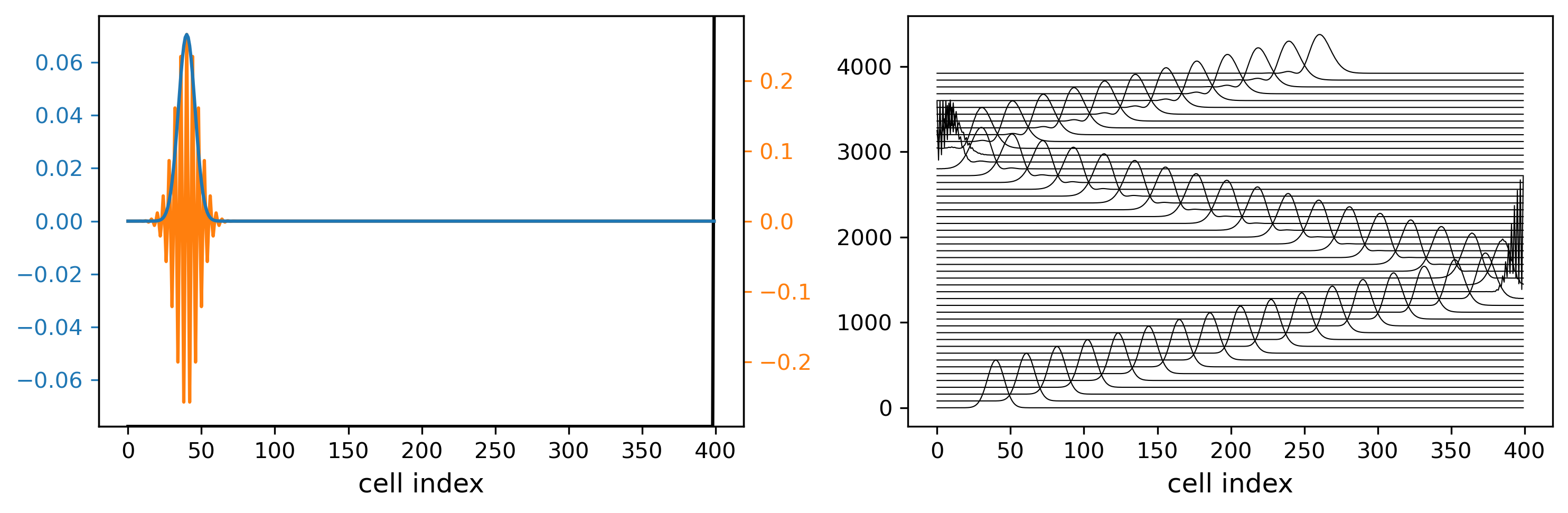}
         \caption{The wave packet is reflected on both sides.}
	\label{fig:box_wavepacket_time}
     \end{subfigure}
     \vspace{3mm}
     
     \begin{subfigure}[b]{\textwidth}
         \centering
    \includegraphics[width=0.999\textwidth]{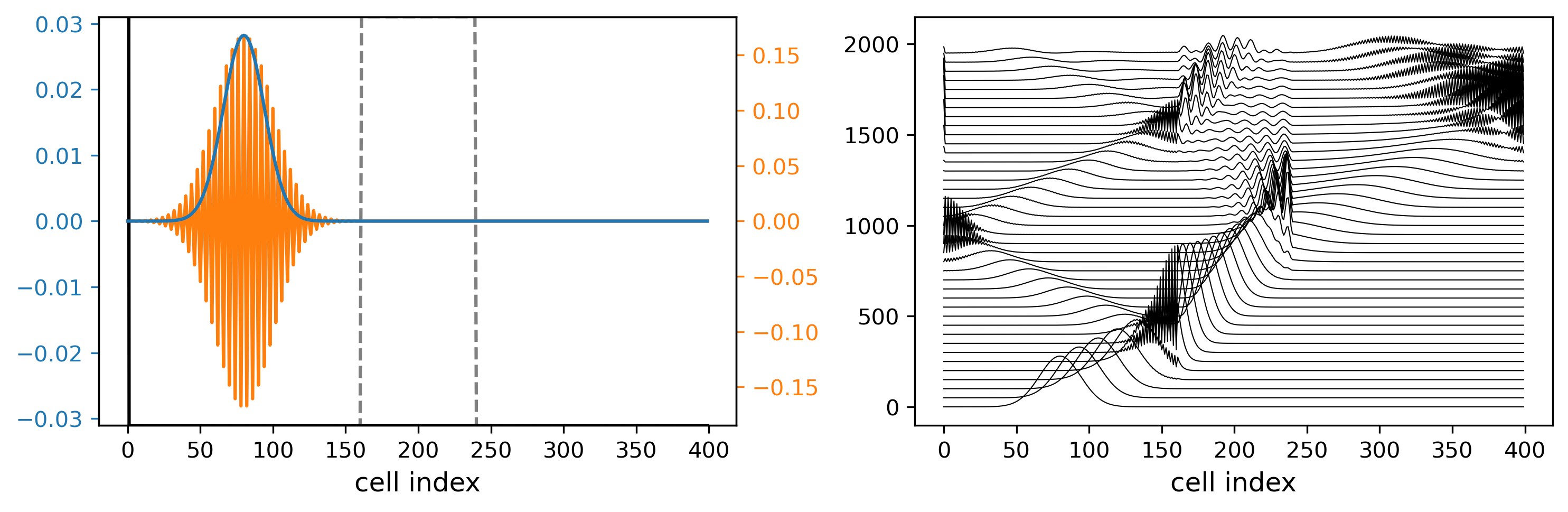}
         \caption{A region with a potential ($V=0.12 \delta_m$) causes refraction.}
	\label{fig:refraction_wavepacket_time}
     \end{subfigure}
\caption{A wave packet in a cellular automaton of 400 cells with reflective boundaries.}
\end{figure}

The cellular automaton of Figure \ref{fig:box_wavepacket_time} has reflective boundaries, 
as in (\ref{eq:reflective}).
The initial wave packet has the same parameters as before:
$\sigma\!=\!8$ and $k \!=\! \frac{\pi}{2}$.
During an evolution of 4000 cycles, 
the wave packet reflects twice at the boundary of the cellular automaton.
With each bounce the wave packet interferes with itself and subsequently recovers.

In addition to reflective boundaries, the cellular automaton of Figure \ref{fig:refraction_wavepacket_time} 
also has a narrow region in the middle with a potential $V(x)=0.24\delta_m$, 
as indicated by the light gray line. 
The automaton is initialized with a wide wave packet and evolves for 2000 cycles.
When the wave packet hits the high-$V(x)$  region, a smaller part is reflected
and the main part proceeds into this region with a reduce group velocity and width.
When this main part hits the backside of the high-$V(x)$ region,
again a smaller part is reflected. 

\subsection{An infinite potential well:}
\label{subsec:infinite_well}

\begin{figure}[!ht]
  \begin{center}
    \includegraphics[width=0.999\textwidth]{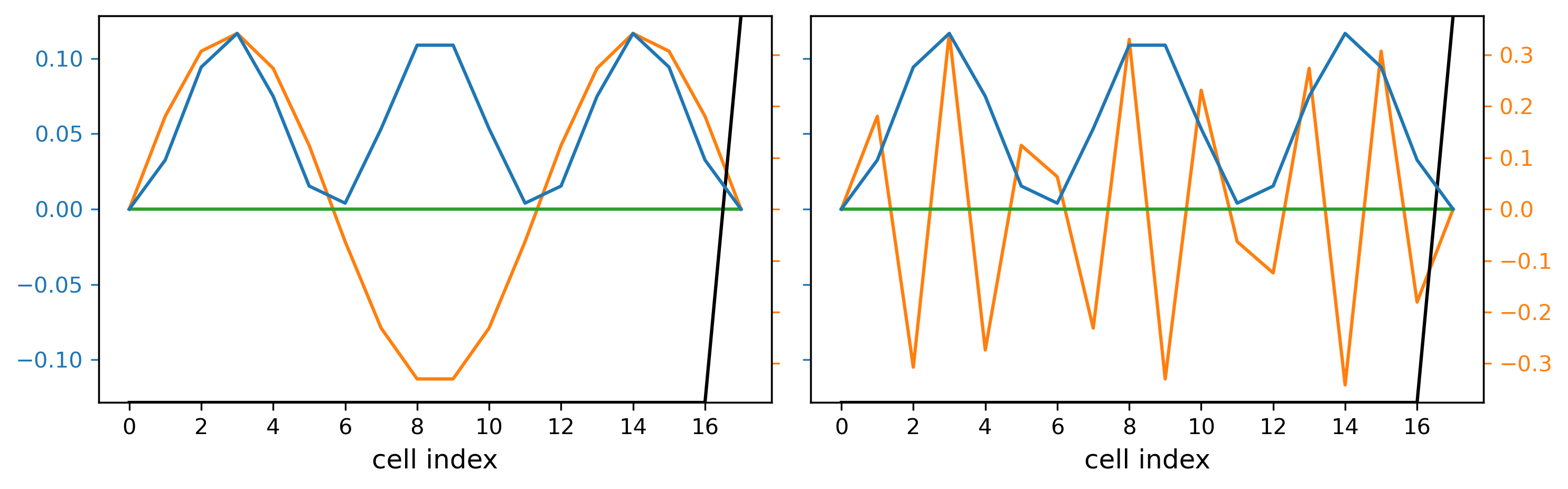}
  \end{center}
  \vspace{-5mm}
\caption{A cellular automaton with an infinite potential well with $L\!=\!17$.
	(left) The initial state is $W_{\mathit box}(\ldots, 3)$.
	(right) The initial state is $W_{\mathit box}(\ldots, L\!-\!3)$.}
\label{fig:box_box_wave}
\end{figure}

The next cellular automaton models an {\em infinite square well} ,
also known as a {\em particle in a box} (see e.g. \cite{2018-Griffiths}).
The box has length $L\!=\!17$ and is centered around $x_c = L/2$.
Inside the box, $x_c \!-\! \frac{L}2 < x < x_c \!+\! \frac{L}2$, the physical potential equals 0, 
outside the box it is infinite, and $\Psi_n(x,t)=0$.
The SCA comprises $L+1$ cells, 
with a reflective cell (\ref{eq:reflective}) at $x=L$ to model the infinite potential.
The initial state is, for integer $n \geq 0$,
\begin{equation} 
W_{\mathit box}(x, x_c, L,n) = \sqrt{\frac2L} \sin \Big( \frac{n\pi}L (x-x_c +L/2) \Big) ~.
  \label{eq:W_box}
\end{equation} 
$W_{\mathit box}$ is subject to spatial aliasing: 
$W_{\mathit box}(x, x_c, L,n) = W_{\mathit box}(x, x_c,L,2L\!+\!n)$,
similar to $W_{\mathit plane}$  (see Subsection \ref{subsec:plane_wave}).
Interestingly, there is another, somewhat more subtle, symmetry.
Let $\Psi_n(t\!=\!0) = W_{\mathit box}(x, x_c,L,n)$ and $\Psi_{L-n}(t\!=\!0) = W_{\mathit box}(x, x_c,L,L\!-\!n)$.
For $t \geq 0$, the corresponding $P(x,t)$ evolve in the same way, that is,
$\sqabs{\Psi_n(x, t)} \;=\; \sqabs{\Psi_{L-n}(x, t)}$.
This symmetry is illustrated in Figure \ref{fig:box_box_wave} for $L\!=\!17$ and $n\!=\!3$.

\begin{figure}[!ht]
  \begin{center}
    \includegraphics[width=0.999\textwidth]{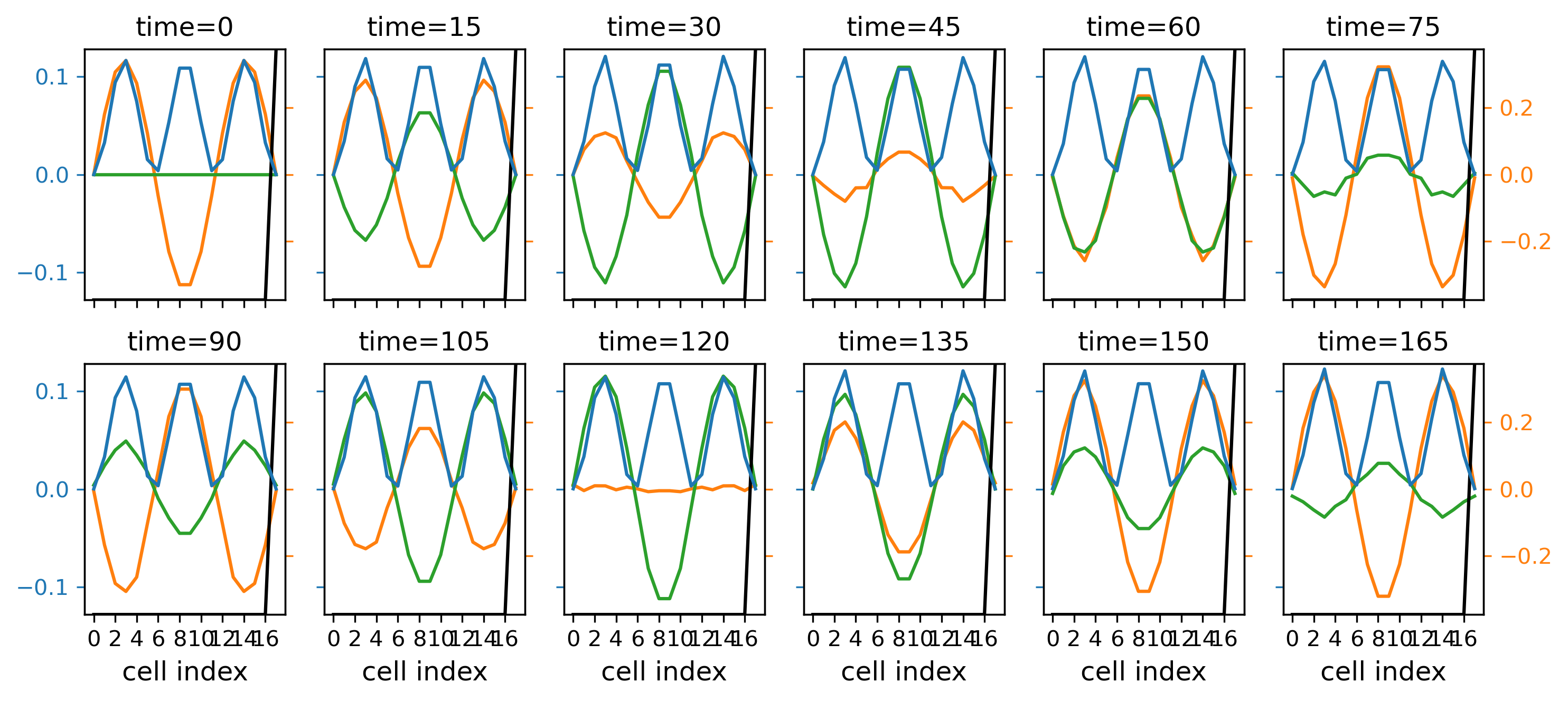}
  \end{center}
  \vspace{-5mm}
\caption{The evolution of the infinite-square-well automaton with $L\!=\!17$.
The colors orange, green, and blue denote 
$\mathrm{Re}(\Psi_3(x,t))$, $\mathrm{Im}(\Psi_3(x,t))$, and $P(\Psi_3(x,t))$ respectively.}
\label{fig:box_box_wave_time}
\end{figure}

\begin{figure}[!ht]
  \begin{center}
    \includegraphics[width=0.999\textwidth]{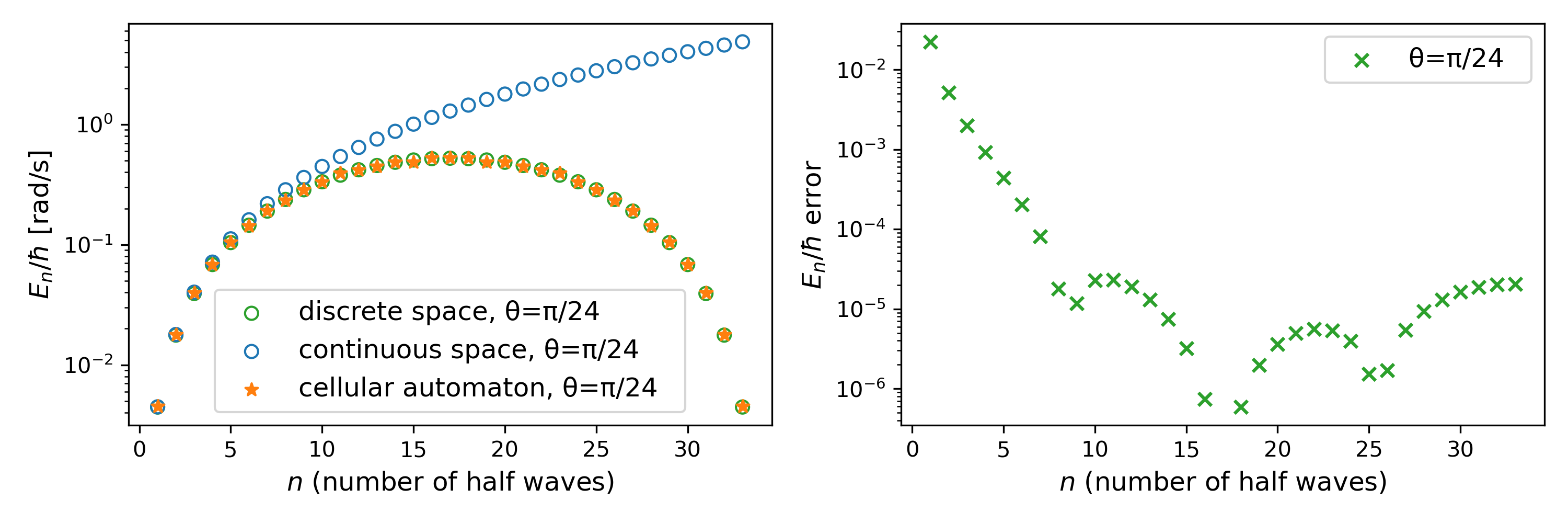}
  \end{center}
  \vspace{-5mm}
\caption{(left) The energy levels $E_n/\hbar$ (in units of $\textrm{rad}/s$) versus $n$ for an infinite square well.
(right) The relative error in energy levels versus $n$.}
\label{fig:well_omega}
\end{figure}

Figure \ref{fig:box_box_wave_time} shows the evolution of the automaton for $n\!=\!3$. 
This evolution is periodic over time with a period of about 160 cycles.
The analytic solution for $\Psi_n(x,t)$ for continuous space is well known, see e.g. \cite{2018-Griffiths}, 
\begin{equation} 
  \Psi_n(x,t) = \sqrt{\frac2L} \sin\Big(k_n \big(x -x_c + \frac{L}2 \big) \Big)
    e^{-i\omega_n t} ~,
  \label{eq:Psi_box}
\end{equation} 
where $k_n = \frac{n\pi}{L}$. The corresponding energy levels are
\begin{equation} 
  \frac{E_n}\hbar
   ~=~ \omega_n 
   ~=~  \frac{n^2 \pi^2 \hbar}{2 m L^2} 
   ~=~  n^2 \pi^2 \Big(\frac{a}{L}\Big)^2 \; \frac\theta\tau 
  ~.
  \label{eq:E_box}
\end{equation} 
Figure \ref{fig:well_omega} (left) shows the matching energy levels,
based dispersion relation (\ref{eq:dispersion_relation}).
The orange stars are derived from the measured oscillation periods
during the evolution of the cellular automaton initialized with $W_{\mathit box}(\ldots ,n)$.
The green circles are based on the dispersion relation.
The blue circles represent (\ref{eq:E_box}).
Figure \ref{fig:well_omega} (right) shows the relative error in energy levels versus $n$,
which is defined analogously to (\ref{eq:v_p-base}) and (\ref{eq:v_p-error}).
Energy $E_n$  is symmetric in $n\!=\!L$ and periodic with period $2L$,
as an immediate consequence of the earlier noted spatial aliasing resulting from the sampling of $W_{\mathit box}$.

\subsection{A harmonic oscillator}
\label{subsec:harmonic_oscillator}

The harmonic oscillator is based on a potential energy function that is parabolic in $x$
\begin{equation} 
	V(\underline{x}) = \frac12 \kappa \omega^2 \underline{x}^2  = \frac12 m \omega^2 \underline{x}^2,
  \label{eq:ho} 
\end{equation} 
where $\kappa$ is the force constant.
Coordinate $\underline{x}$ is underlined to distinguish it from cell index $x$.
The normalized stationary states for the harmonic oscillator are  (e.g. \cite{2018-Griffiths}):
\begin{equation} 
	\psi_n(\underline{x}) = \left( \dfrac{m\omega}{\pi\hbar} \right)^{1/4}
               		 \dfrac{1}{\sqrt{2^n n!}} H_n (\xi) \exp{\left( -\xi^2/2 \right)}
	\hspace{1cm} \text{with} \hspace{1cm}
	 \xi = \sqrt{\dfrac{m\omega}{\hbar}} \underline{x} ~,
  \label{eq:ho_stationary}
\end{equation} 
where $n$ is the state number, $\xi$ is a dimensionless variable, 
and $H_n (\xi)$ is the $n$th-order Hermite polynomial.
For the experiment, state $n\!=\!3$ is chosen.
In order for the  $x$-extent of the waveform to cover a significant portion of the $N \!=\! 400$ cells,
$x$ is scaled according to $x = \rho \xi$.
The initial state then becomes
\begin{equation} 
  W_{\mathit harmonic}(x,n) = \psi_n( \rho \xi) ~,
  \label{eq:W_harmonic} 
\end{equation} 
with $n\!=\!3$ and  $\rho\!=\!N/12$ for the experiment. 
The potential $V(x) = \frac12 \kappa (x\!-\!x_c)^2$ requires a value for $\kappa$.
This value is obtained by  iteratively tuning $\kappa$ towards a stationary state for $n\!=\!3$, 
with  $\kappa \!=\!  2.18 \cdot 10^{-7} \delta_m$ as outcome.

\begin{figure}[!ht]
  \begin{center}
    \includegraphics[width=0.999\textwidth]{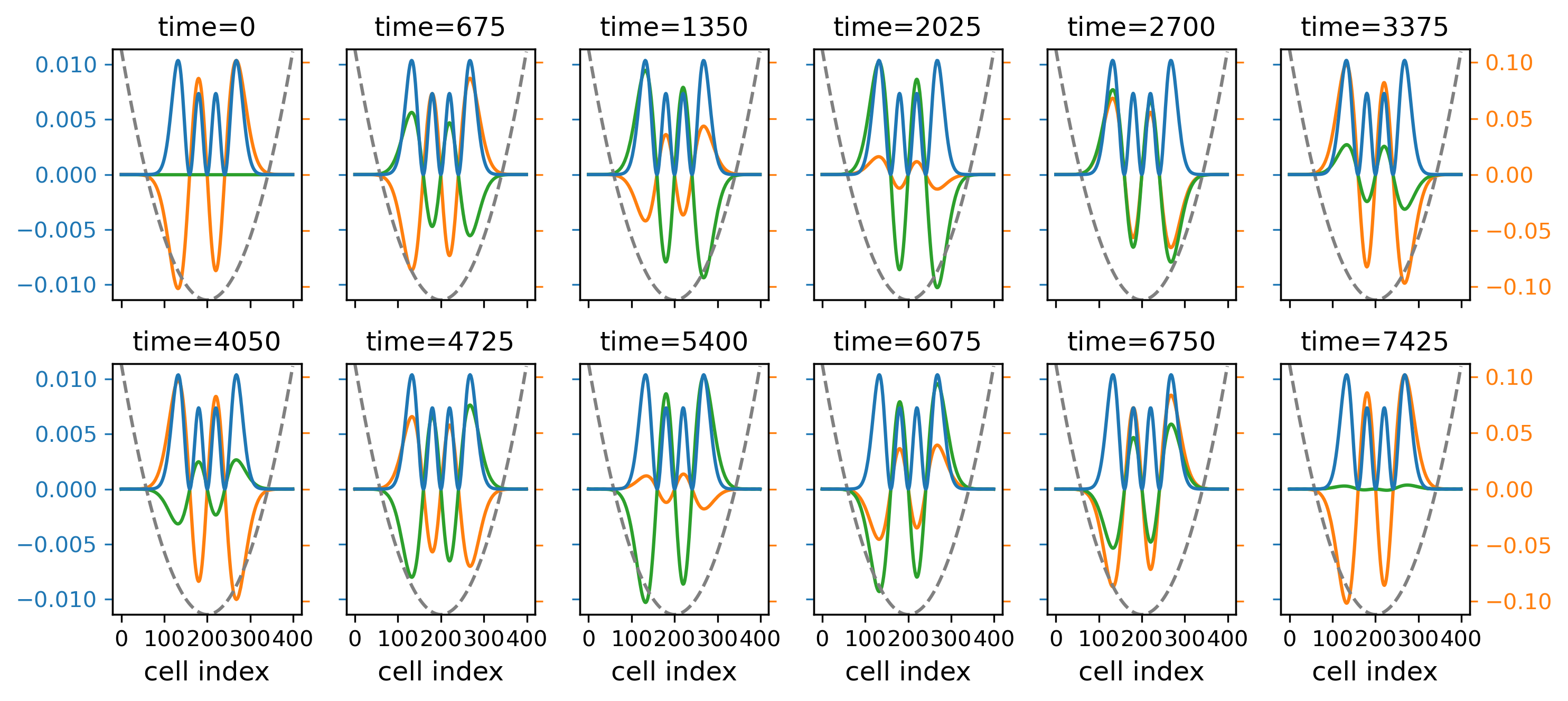}
  \end{center}
  \vspace{-5mm}
\caption{The evolution of a harmonic-oscillator CA initialized with $W_{\mathit harmonic}(x,3)$.
		The dashed parabola represent $V(x)$. The measured oscillation period is 7425 cycles.}
\label{fig:oscillator_W_harmonic}
\end{figure}

Figure \ref{fig:oscillator_W_harmonic}  shows the evolution of the cellular automaton, 
where state $\Psi(x,t)$ is captured every 675 cycles.
A single oscillation period lasts about 7425 cycles,
which corresponds to $\omega_3 = 2\pi /7425 \approx 0.00085$ radians per second.
Probability $P(x,t)$ (in blue) is constant over the entire episode.

For continuous space $\omega_n = (n+\frac12 \omega)$.
The value for  $\omega$ can be derived from $x = \rho \xi \underline{x}~$.
With cell size $a$, also $\underline{x} = a x$ must hold. Hence,
\begin{equation} 
	\rho \xi ~=~\rho \sqrt{\dfrac{m\omega}\hbar} ~=~1
	\hspace{1cm} \text{and} \hspace{1cm} 
	\omega = \frac\hbar{m\rho^2} = \frac{2 \theta}{\rho^2}~.
  \label{eq:ho_energy}
\end{equation}

\begin{figure}[!ht]
  \begin{center}
    \includegraphics[width=0.999\textwidth]{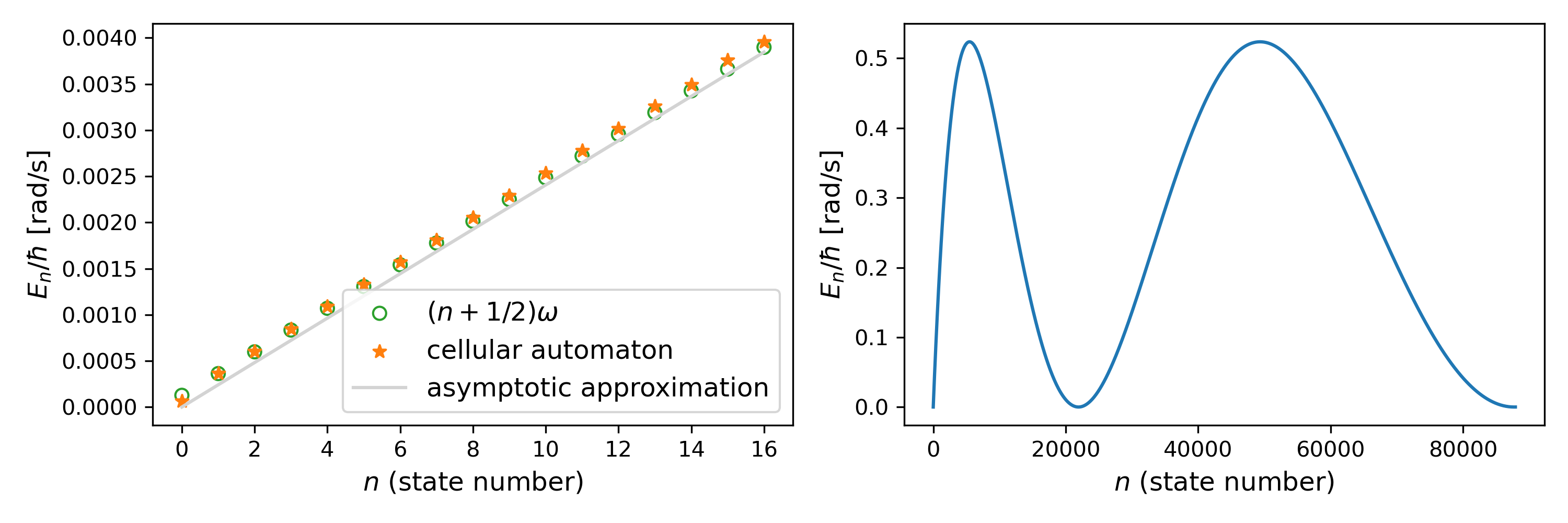}
  \end{center}
  \vspace{-5mm}
\caption{
(left) The energy levels $E_n/\hbar$ of the harmonic oscillator for $n\leq 16$.
	The orange $*$s show the measured frequencies $\omega_n$ of $\Psi_n(x,t)$ and
	the green circles show $(n+\frac12)\omega$.
(right) The energy levels for large $n$, based on an asymptotic expansion of $H_n (\rho \xi)$.}
\label{fig:harmonic_energy}
\end{figure}

Figure \ref{fig:harmonic_energy} (left) shows the energy levels $E_n/\hbar = \omega_n$ for $n \!\leq\! 16$.
Each orange $*$ is obtained by initializing the cellular automaton with $\psi_n$,
and evolving the automaton until it returns to $\psi_n$. 
The energy level $\omega_n$ equal the reciprocal values of the measured time intervals.
The thus  obtained  $\omega_n$ agree well with  $\omega_n = (n+\frac12)\hbar\omega$.

To get an impression for the CSA behavior for large $n$,
an asymptotic expansion of the Hermite polynomials can be used,
e.g. based on \cite{1965-Abramowitz}, formula 13.6.38, parameters 13.5.16.
Asymptotically, for $n \rightarrow \infty$, this expansion is of the form
\begin{equation} 
  A(n) \cos \big(\underline{x} \sqrt{2n} - \frac{n\pi}2\big) B(n,\underline{x}) ~,
  \label{eq:Hermite_expansion}
\end{equation} 
where $A(n)$ does not depend on $\underline{x}$ and $B(n,\underline{x})$ has only two zeros in $\underline{x}$.
This expansion suggests an ``effective wave number'' $k = \frac1\rho \sqrt{2n}$.
It also implies that at the bottom of the $V(\underline{x})$ parabola 
there is a half cosine  with $V(\underline{x}) \approx 0$.
Application of the dispersion relation {\em of the free particle}\footnote{
   While this ignores the potential $V(x)$ of the oscillator,
   it does highlight the effects of spatial aliasing.}
(\ref{eq:dispersion_relation}) to the ``effective'' $k$ cited above results in   
\begin{equation} 
	\omega_n (\theta, \rho)
	   ~\approx~   4 \theta \sin^2{\left(\frac1{\rho} \sqrt{\frac{n}2} \right)} 
       			+ \frac43 \theta^{3} \cos{\left( \frac1{\rho}\sqrt{2n}\right)}\sin^2{\left(\frac1{\rho}\sqrt{2n}\right)}
    			+ O\left(\theta^{4}\right) ~.
  \label{eq:harmonic_asymptotic}
\end{equation} 

Figure \ref{fig:harmonic_energy} (right) shows $\omega_n (\theta, \rho)$ versus $n$
for $\theta \!=\! \pi/24$ and $\rho \!=\! 400/12$.
Its quasi periodic dependency on state number $n$ is, again, 
an immediate consequence of the spatial aliasing resulting from the sampling of $W_{\mathit harmonic}$.
For $n \!\leq\! 16$ these asymptotic levels are also depicted in gray in the (left) panel,
where they agree well with the measured $\omega_n$ .

\subsection{A time-dependent harmonic oscillator}
\label{subsec:adiabiatic}

\begin{figure}[!h]
     \centering
     \begin{subfigure}[b]{\textwidth}
         \centering
    \includegraphics[width=0.999\textwidth]{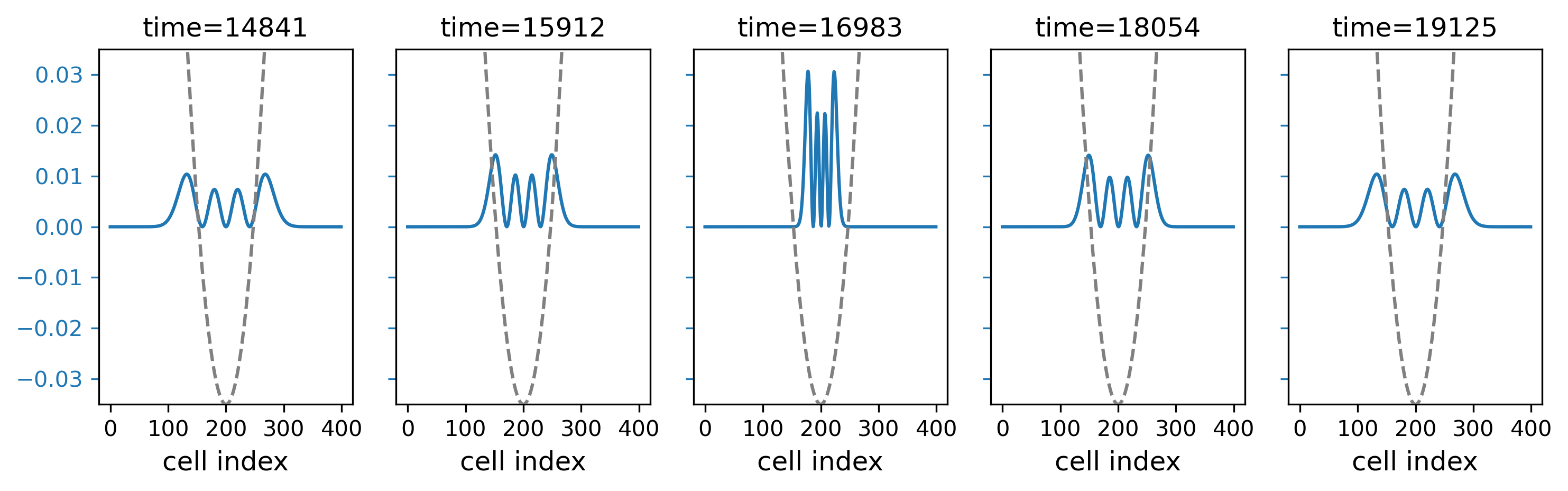}
         \caption{After an instantaneous increase the waveform quasi-periodically narrows and widens.}
	\label{fig:ho-adiabatic}
     \end{subfigure}
     \vspace{3mm}
     
     \begin{subfigure}[b]{\textwidth}
         \centering
    \includegraphics[width=0.999\textwidth]{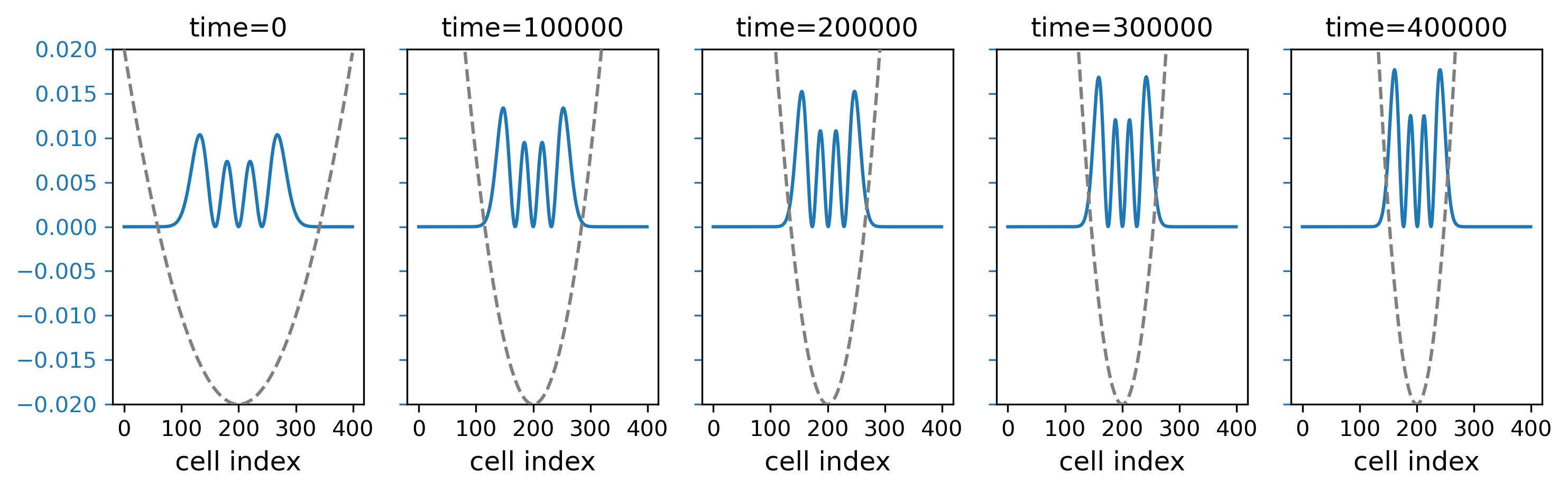}
         \caption{During an adabatic increase over 400,000 cycles the eigenstate gradually narrows.}
	\label{fig:ho-squeeze}
     \end{subfigure}
\caption{The evolution of the harmonic oscilator for the $n\!=\!3$ eigen state in response to  
ninefold increase of the potential. (Note the difference in time frames.)}
\label{fig:adiabatic-time}
\end{figure}

The harmonic oscillator of the previous subsection becomes time dependent 
with a time-dependent potential $V(x,t)$, as e.g. in
\begin{equation} 
   V(x,t) =  s(t) V(x)  \hspace{2cm}  s(t) =  \min(\max(1 +  \alpha (t -t_s), \;0), \;s_{max}) ~.
    \label{eq:slope}
\end{equation} 
Slope $s(t)$ is 0 initially, and from $t_s$ onwards increases by $\alpha$ per iteration
until  $s_{max}=9$ is reached.
In the experiment of Figure \ref{fig:adiabatic-time} the harmonic oscillator 
is initialized with the same state as in Figure \ref{fig:oscillator_W_harmonic}
and $V(x,t)$ is updated every ten iterations.

\begin{figure}[!ht]
  \begin{center}
    \includegraphics[width=0.999\textwidth]{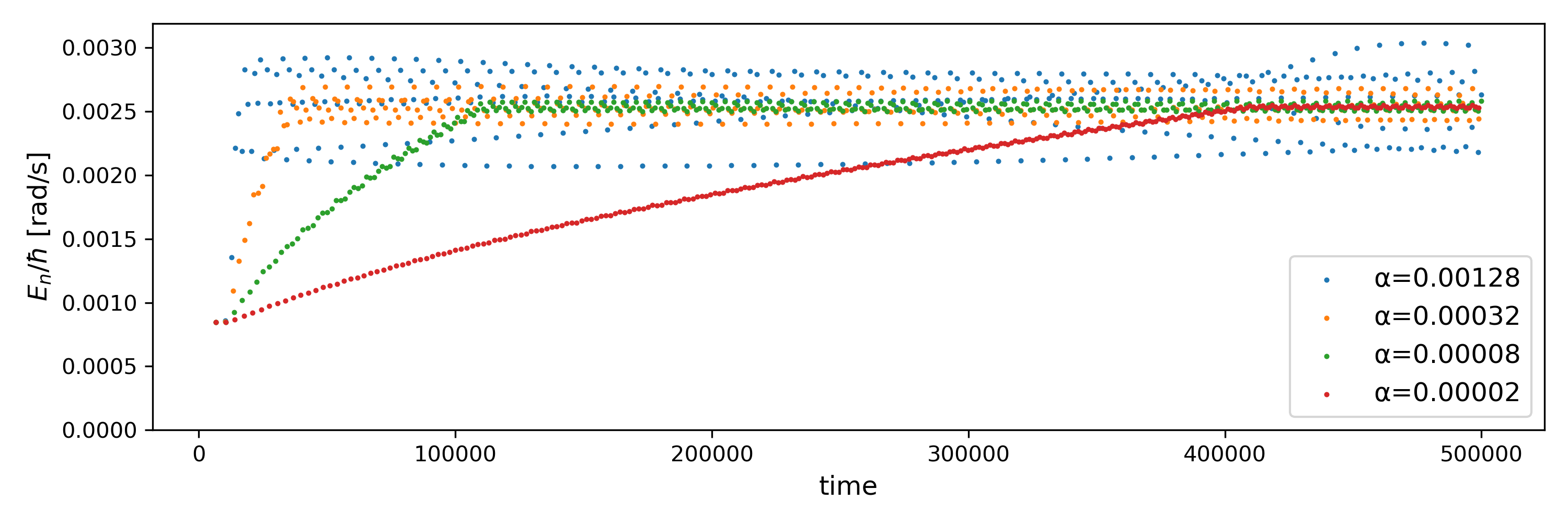}
  \end{center}
  \vspace{-5mm}
\caption{
   The energy levels $E_n/\hbar$ of the harmonic oscillator for $n\!=\!3$ with $V(x,t)$ increasing over time.
   The red trajectory is marginally adiabatic; the eigen state evolves gradually. 
   The green, orange, and blue ones are clearly diabatic, leading to squeezed states.
}
\label{fig:adiabatic}
\end{figure}

Figure \ref{fig:adiabatic} shows the energy levels $E_3 (t)/\hbar = \omega(t)$ for 
four values of $\alpha$, ranging from 0.00002 to 0.00128. 
The colored markers are derived from $I(t)$,
\begin{equation} 
  I(t) = \sum_x \left| \text{Im}(\Psi (x,t)) \right| ~.
  \label{eq:omega_t}
\end{equation} 
For a time-independent Hamiltonian $I(t) = \left| \sin (\omega t+\phi) \right|$
for some $\phi$.
Then curve $I(t)$ touches zero periodically, with a fixed time interval $\pi/\omega$.
For a time-dependent Hamiltonian, 
the time intervals $t_i$ between successive minima of $I(t)$ vary over time. 
The marker series of Figure \ref{fig:adiabatic}, per $\alpha$ value, 
represent the series $(T_i, \omega_i)$, where $\omega_i= \pi/t_i$ and $T_i = \sum_{j\leq i} t_j$.

\begin{figure}[!ht]
  \begin{center}
    \includegraphics[width=0.999\textwidth]{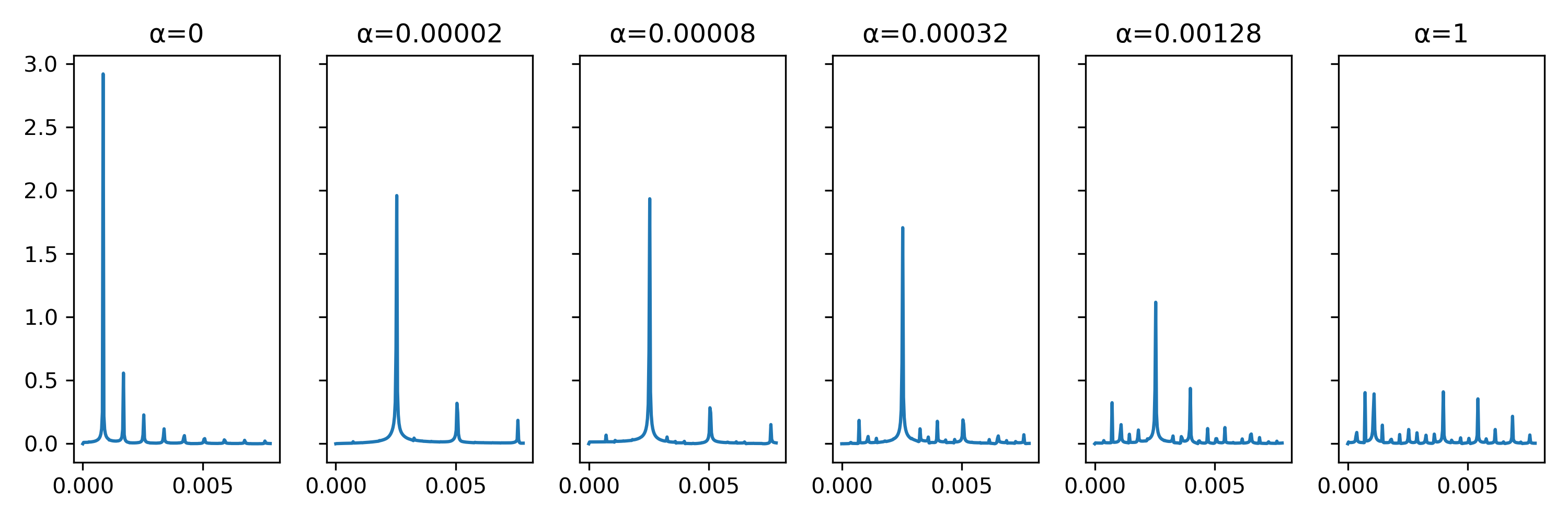}
  \end{center}
  \vspace{-5mm}
\caption{
  Frequency spectra of $I(t)$ for time interval [400k,500k]  for various values of $\alpha$.
}
\label{fig:ffts}
\end{figure}

The time-dependent harmonic oscillator has been analyzed extensively,
as, for example, in \cite{1994-Kiss, 2021-Martínez},
mostly with an emphasis on theory.
Schr{\" o}dinger cellular automata may offer a new tool to their study.
For example, Figure \ref{fig:ffts} shows 
a low-frequent portion of the FFT transform of $I(t=[400k,500k])$ 
for various values of $\alpha$, after elimination of the DC component.
For $\alpha\!=\!0$, the evolution is time independent, 
with a fundamental frequency $\omega \!=\! 0.0008$ radians per second.
This is consistent with the observed 7425 cycles in Figure \ref{fig:oscillator_W_harmonic}.
The higher harmonics are a byproduct of the absolute operation in de definition of $I(t)$ in
(\ref{eq:omega_t}).
For the adiabatic evolution ($\alpha\!=\!0.00002$), the entire spectrum is essentially stretched by $3\times$, as to be expected.
For higher values of $\alpha$, the spectrum becomes increasingly polluted 
by components that are both higher {\em and lower} in frequency than the fundamental one.
The spectrum for $\alpha\!=\!1$ is rather messy; the fundamental frequency is barely visible, 
and a proper detection of the minima of $I(t)$ appears infeasible.

\section{Experiments with 2D Schr{\" o}dinger cellular automata}
\label{sec:2D-experiments}

This section describes five experiments with 2D SCA.
All SCA apply the same $\theta = \pi/24$, 
but differ in scalar potential $V(\xvec)$,  reflective surfaces $X(\xvec)$, and/or vector potential $\AAA(\xvec)$.
The results are visualized as 2D color plots of $\sqabs{\Psi}$
based  on a {\em qualitative} colormap, in order to highlight the geometric features of the wave forms.
The light blue backdrop implies a negligible $\sqabs{\Psi}$
and white indicates a high value of $\sqabs{\Psi}$.

\newpage

\subsection{Refraction}
\label{subsec:refraction}

\begin{figure}[!h]
     \centering
     \begin{subfigure}[b]{\textwidth}
         \centering
         \includegraphics[width=0.95\textwidth]{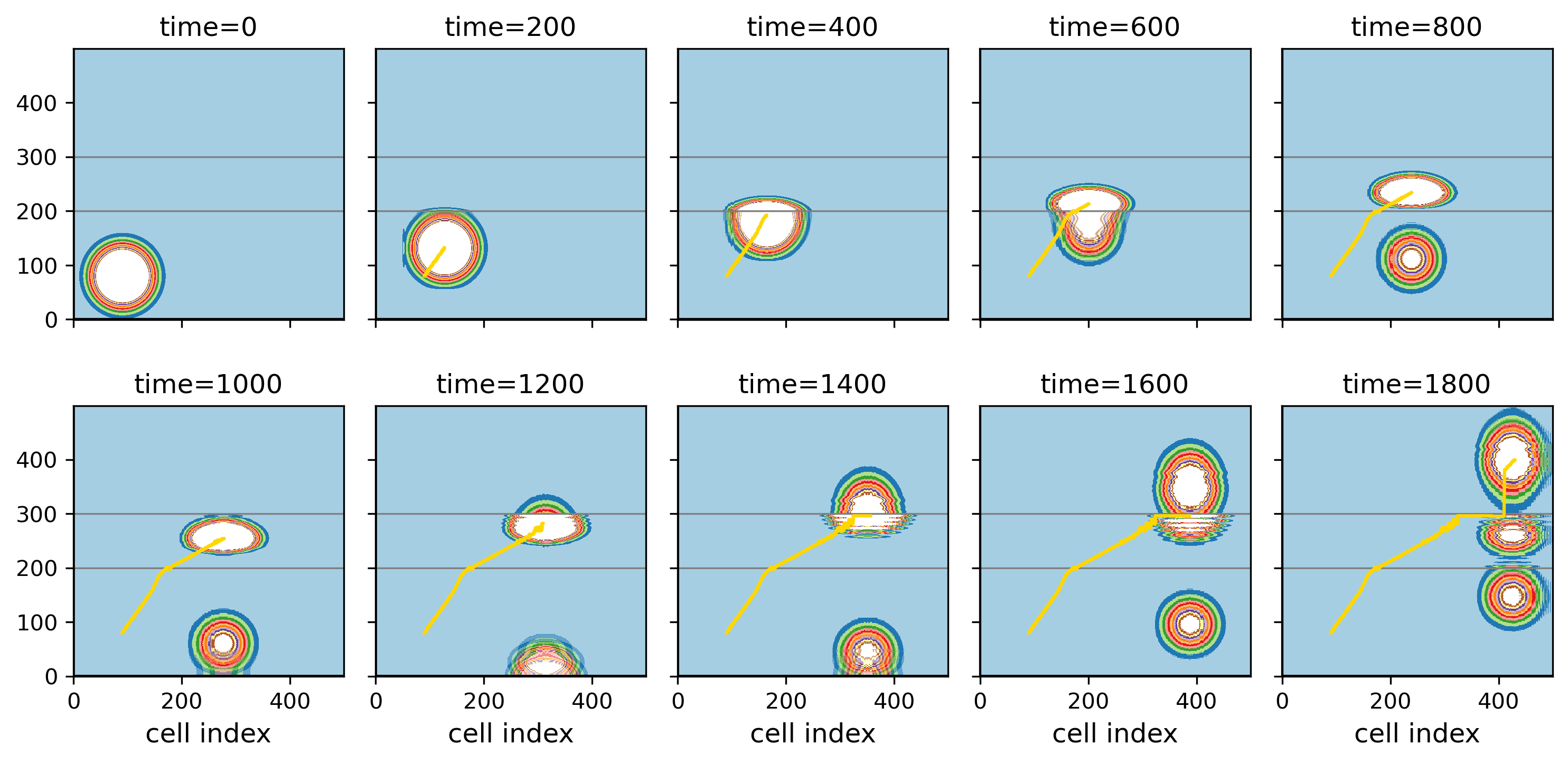}
         \caption{The region in the middle has a fixed potential $V_{rf} = 0.24 \delta$.}
         \label{fig:refraction-V-region}
     \end{subfigure}
     \vspace{3mm}
     
     \begin{subfigure}[b]{\textwidth}
         \centering
         \includegraphics[width=0.95\textwidth]{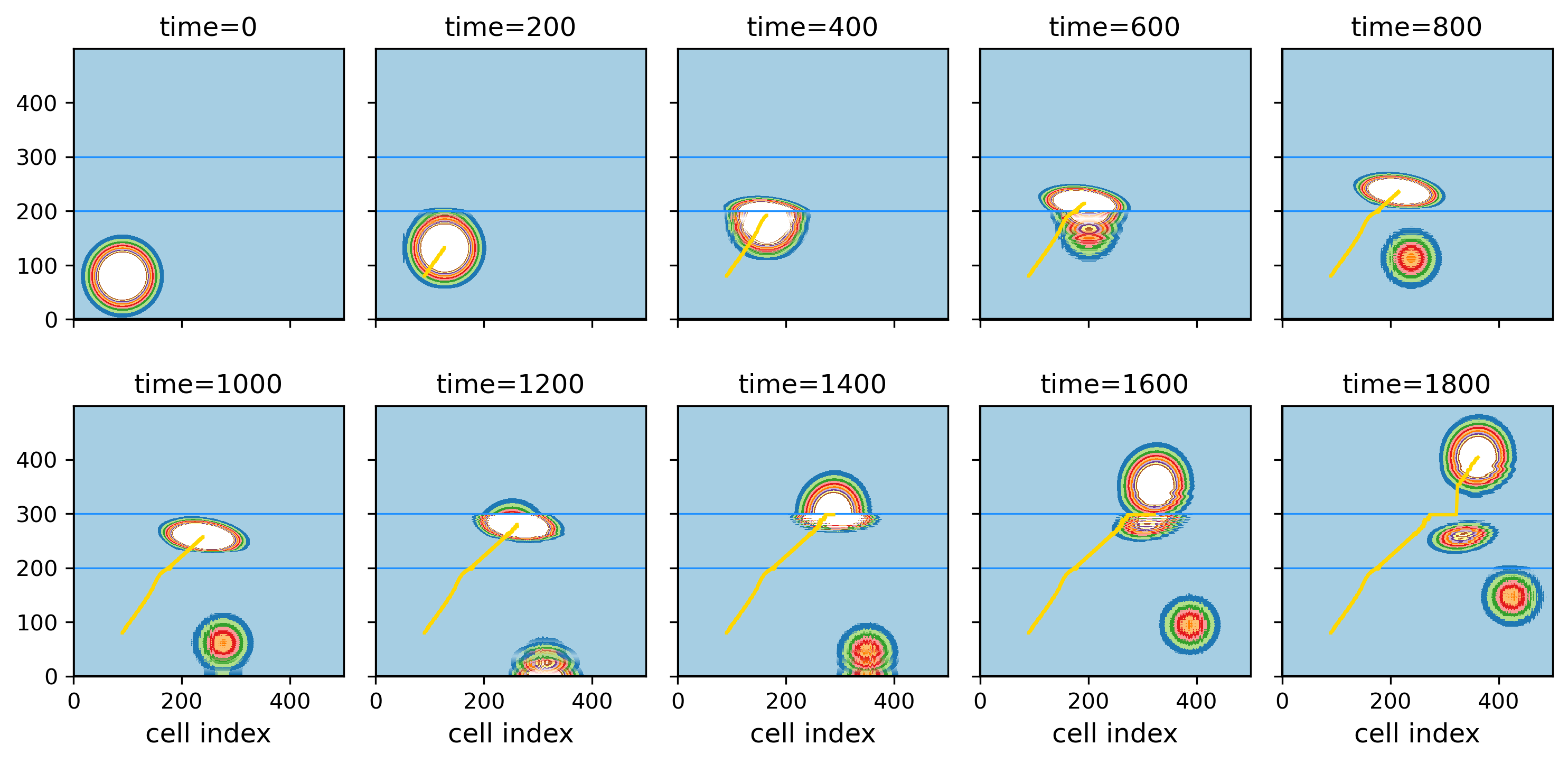}
         \caption{The region in the middle has a reduced value for $\theta$:  $\theta_{rf} \!=\!\pi/38$ .}
         \label{fig:refraction-theta-region}
     \end{subfigure}
\caption{Two refraction experiments of a 2D Gaussian wave packet.}
\label{fig:refraction}
\end{figure}

Refraction is the redirection of a wave as it passes from one medium to another.
The refraction experiment of Figure \ref{fig:refraction-V-region} is a 2D version 
of the 1D refraction automaton of Figure \ref{fig:refraction_wavepacket_time}.
The horizontal region in the middle has a potential $V_{rf} = 0.24 \delta$.
The 2D wave packet partially bounces off that region 
and partially propagates through that region at a reduced velocity. 
Subsequently, the reduced packet further splits into two parts, 
one escaping from the $V$ region, and the other remaining inside.
Note that the vertical component of the wave-packet velocity is affected, 
unlike the horizontal one.

The inhomogeneous SCA of Figure \ref{fig:refraction-theta-region} is similar,
but now the middle region has a reduced value for $\theta$: $\theta_{rf} \!=\! \pi/38$.
This results in a reduced group velocity  in both $x$ and $y$.
Neither refraction experiment is consistent with Snell's law of refraction.


\subsection{The Davisson-Germer experiment}
\label{subsec:Davisson-Germer}

The Davisson-Germer experiment \cite{1927-Davisson} in 1927 confirmed the wave-particle duality hypothesis
(de Broglie), and thereby the wave mechanics approach of the Schr{\" o}dinger equation.

\begin{figure}[!h]
     \centering
     \begin{subfigure}[b]{\textwidth}
         \centering
         \includegraphics[width=0.95\textwidth]{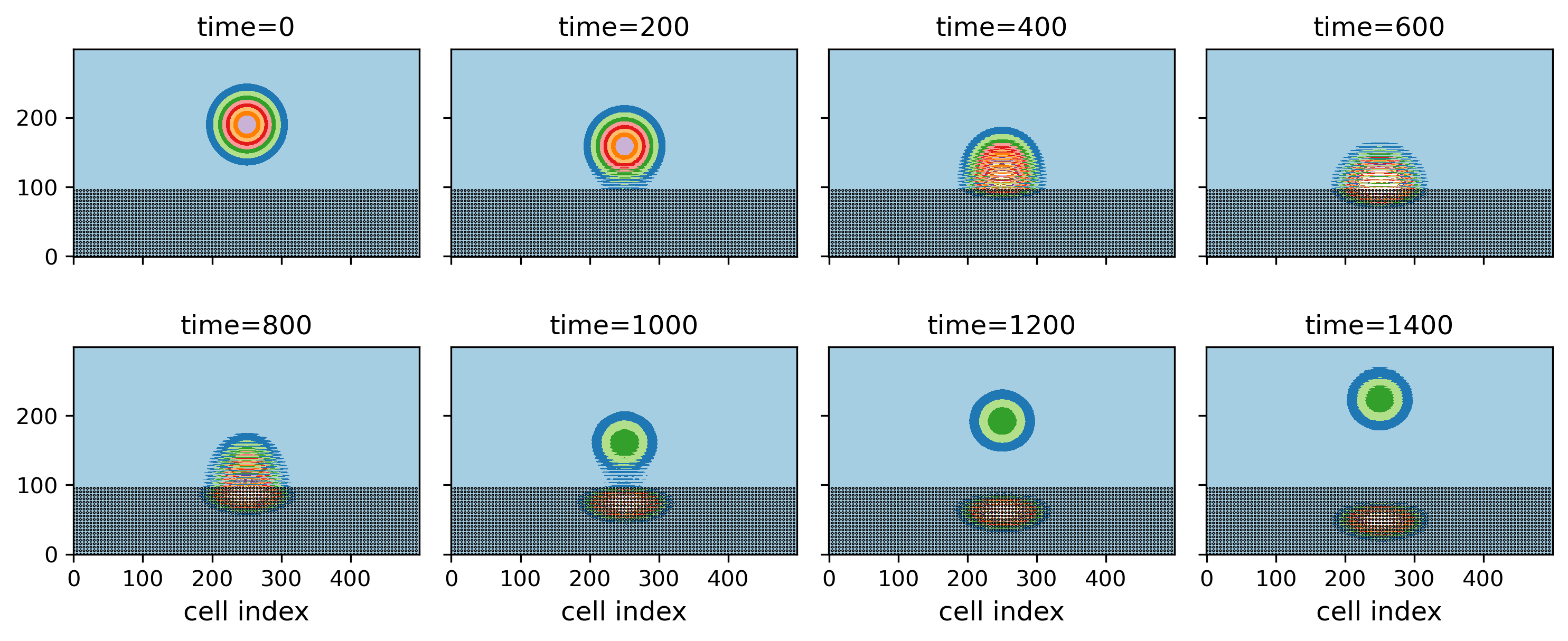}
         \caption{With an angle of incidence of $90^o$ the wave packet is mostly absorbed.}
         \label{fig:Davisson-Germer-90}
     \end{subfigure}
     \vspace{3mm}
     
     \begin{subfigure}[b]{\textwidth}
         \centering
         \includegraphics[width=0.95\textwidth]{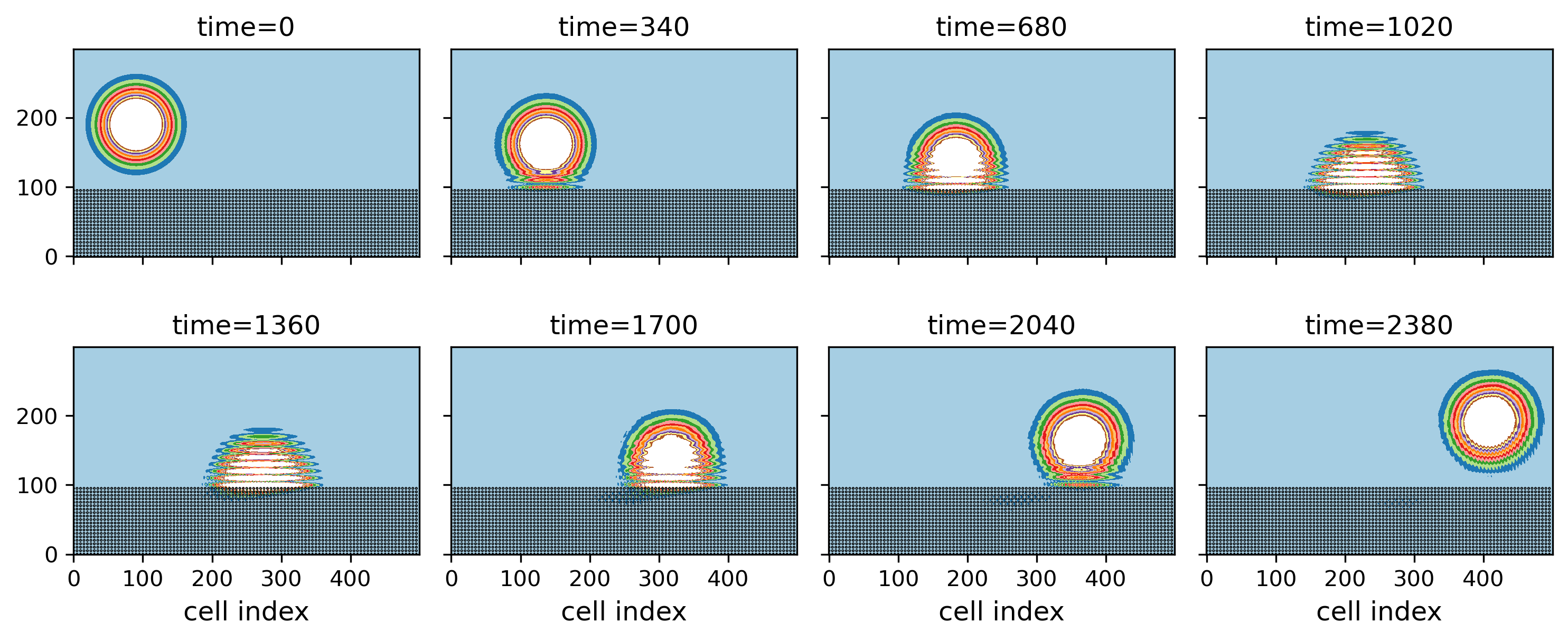}
         \caption{With an angle of incidence of $30^o$ the wave packet is almost entirely reflected.}
         \label{fig:Davisson-Germer-30}
     \end{subfigure}
     \caption{The Davisson-Germer experiment.}
     \label{fig:Davisson-Germer}
\end{figure}

In Figure \ref{fig:Davisson-Germer} the surface of the original crystal of nickel metal
is modelled by a regular grid of single-cell reflectors (Subsection \ref{subsec:X}),
spaced by five cells (half a wavelength) in each direction.
Davisson and Germer fired slow-moving electrons at this crystalline nickel target
and observed a diffraction pattern,
where the reflected electron intensity depends on the angle of incidence.
This diffraction can be fully explained in terms of interference, using Bragg's law.
The SCA of Figure \ref{fig:Davisson-Germer} visualizes such differences in diffraction
for 90 and 30 degree angles of incidence of a wave packet.


\subsection{A Mach-Zehnder interferometer}
\label{subsec:Mach-Zehnder}

A Mach-Zehnder (MZ) interferometer is an instrument to measure phase-shift variations
and was proposed by Ludwig Mach and Ludwig Zehnder in the late 19th century.
Practical instruments are based on photons (collimated light),
but early this century Ji et al \cite{2003-Ji} demonstrated a Mach-Zehnder interferometer based on electrons.

\begin{figure}[!h]
     \centering
     \begin{subfigure}[b]{\textwidth}
         \centering
         \includegraphics[width=0.95\textwidth]{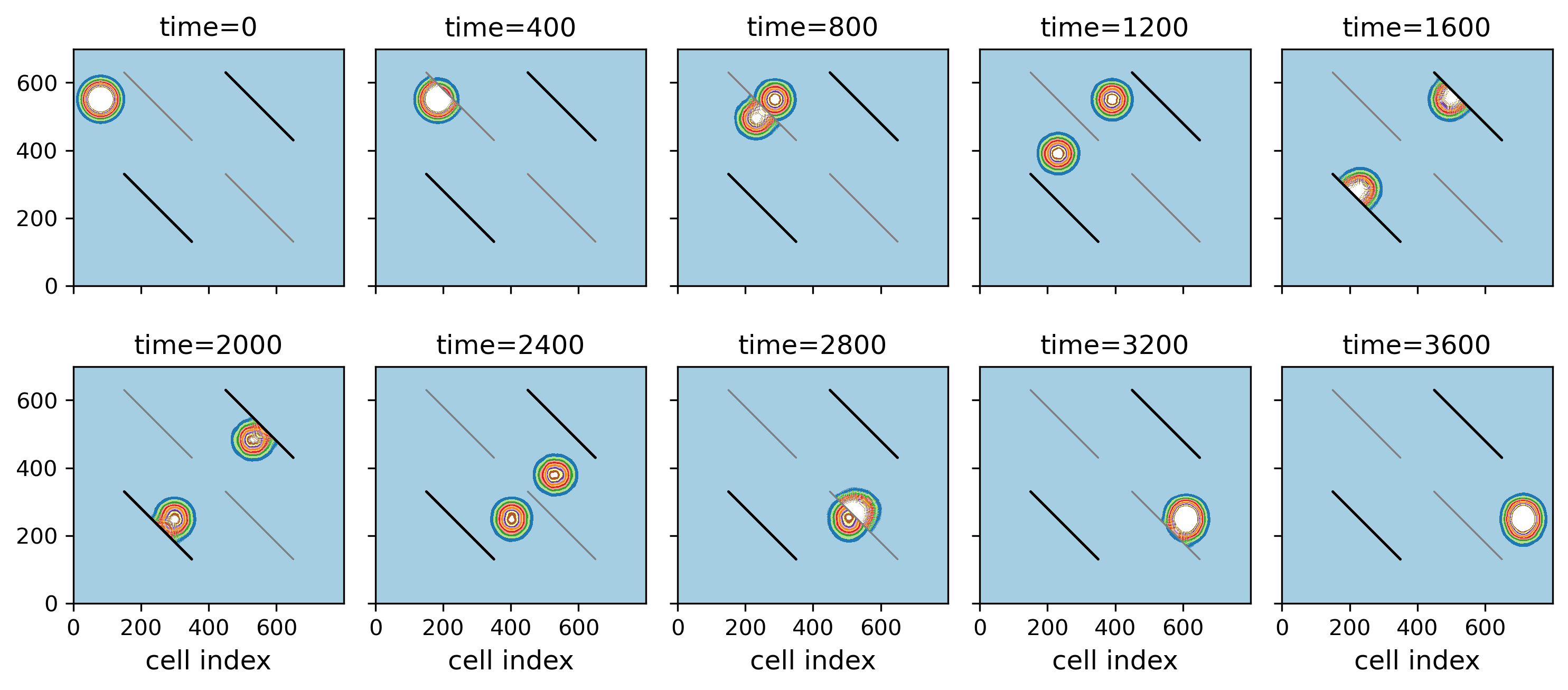}
         \caption{The lengths of the top-right and left-bottom paths are equal.}
         \label{fig:Mach-Zehnder-symmetric}
     \end{subfigure}
     \vspace{3mm}
     
     \begin{subfigure}[b]{\textwidth}
         \centering
         \includegraphics[width=0.95\textwidth]{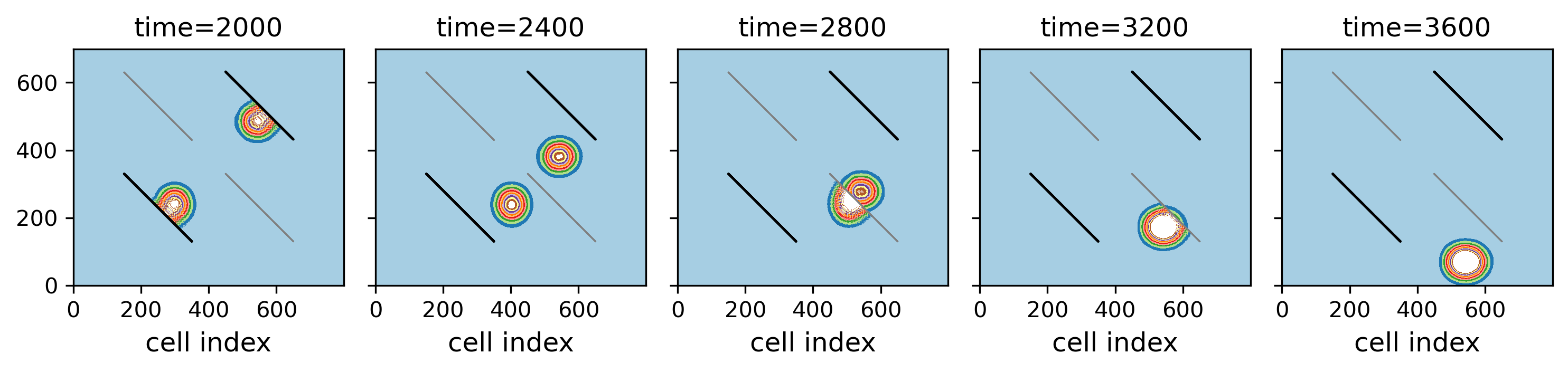}
         \caption{The lengths of the top-right and left-bottom paths differ by half a wavelength.}
         \label{fig:Mach-Zehnder-asymmetric}
     \end{subfigure}
     \caption{The Mach-Zehnder interferometer.}
     \label{fig:Mach-Zehnder}
\end{figure}

The prototypical MZ interferometer of Figure \ref{fig:Mach-Zehnder} 
consists of two mirrors (black) and two beamsplitters (gray), all placed under a 45 degree angles.
The mirrors are reflective single-cell-width surfaces (Subsection \ref{subsec:X}),
and the beam splitters are single-cell-width surfaces with a potential $V_{bs} = 0.248 \delta$.
The difference in outcomes (wave-packet exit directions)
for Figures \ref{fig:Mach-Zehnder-symmetric} and \ref{fig:Mach-Zehnder-asymmetric} 
stems from the respective path-lengths differences.
The SCA evolutions nicely visualize that the MZ interferometer must be described by a genuine quantum superposition of the two wave-packet paths.


\subsection{The double-slit experiment}
\label{subsec:double-slit}

The double-slit experiment demonstrates quantum interference of a single massive particle.
Feynman describes this form
of interference as {\it  ``a phenomenon which is impossible, absolutely impossible, 
to explain in any classical way, and which has in it the heart of quantum mechanics''} \cite{2009-Feynman-Vol3}.
Single-electron interference, using a biprism beam splitter, was demonstrated by Merli et al  \cite{1976-Merli}.

\begin{figure}[!h]
     \centering
     \begin{subfigure}[b]{\textwidth}
         \centering
         \includegraphics[width=0.95\textwidth]{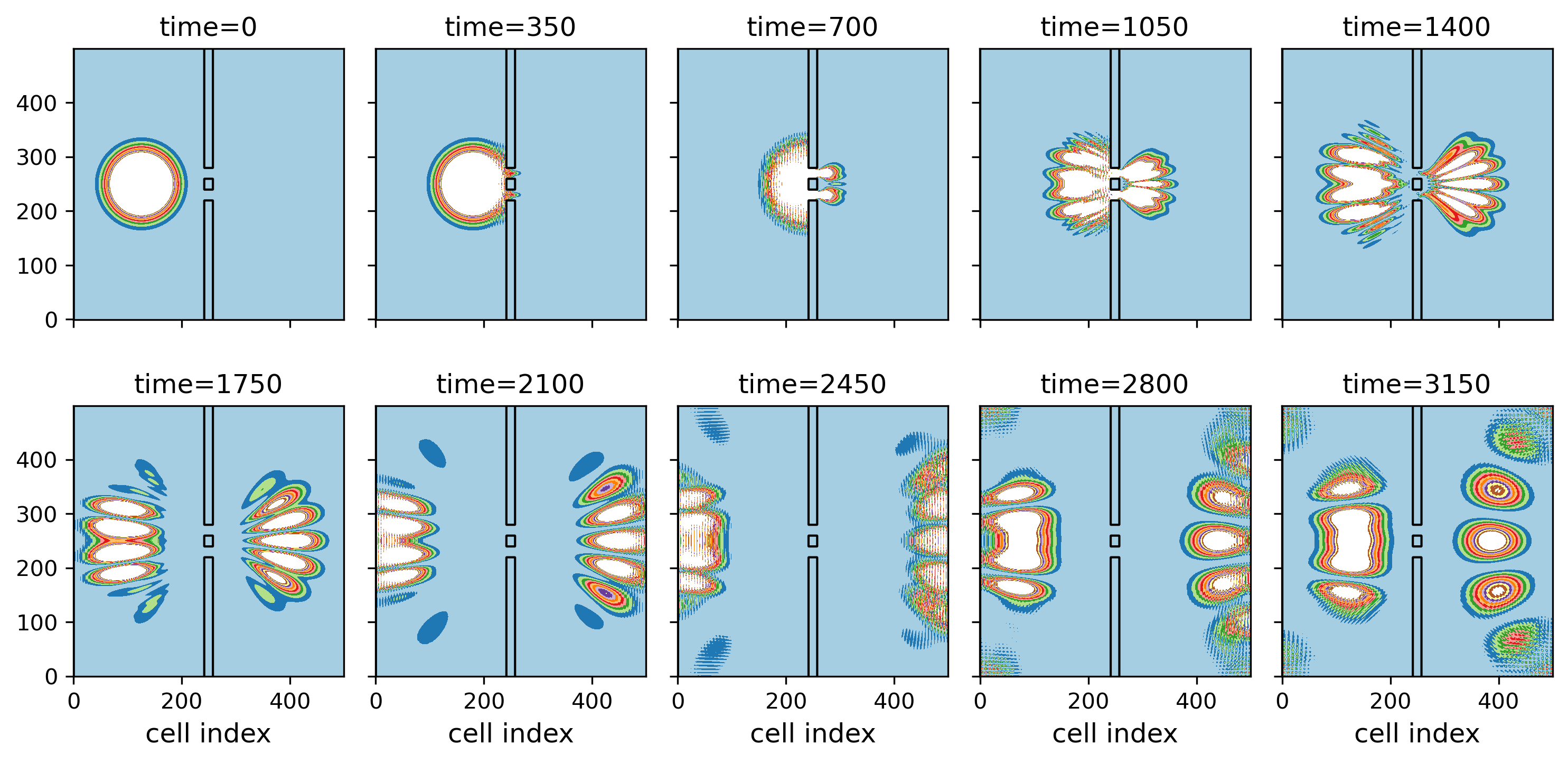}
         \caption{Colors show different values of $\sqabs{\Psi}$.}
         \label{fig:double-slit-P}
     \end{subfigure}
     \vspace{3mm}
     
     \begin{subfigure}[b]{\textwidth}
         \centering
          \includegraphics[width=0.95\textwidth]{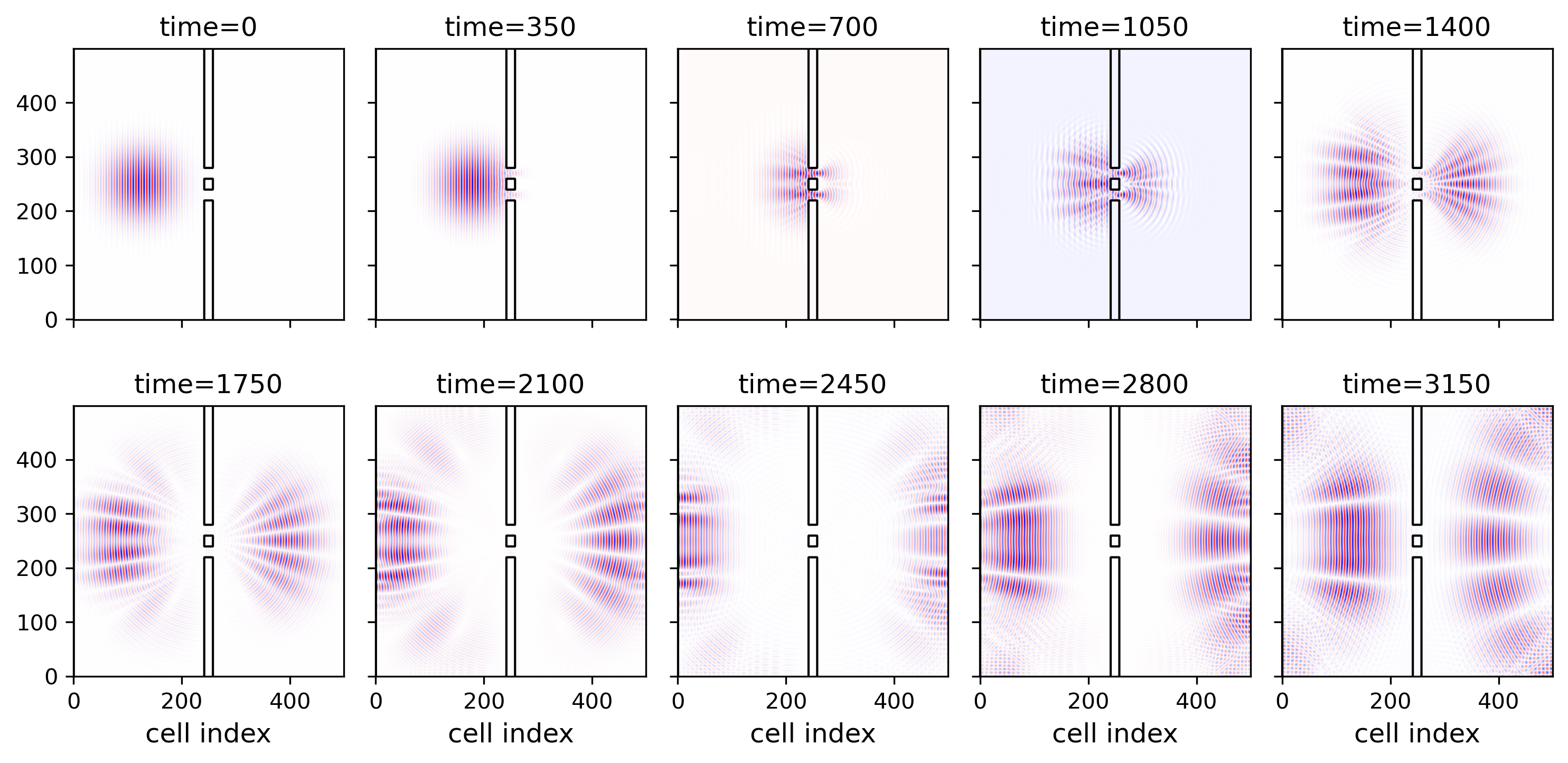}
         \caption{The color red (blue) shows positive (negative) values of $\textrm{Re}(\Psi)$.}
         \label{fig:double-slit-real}
     \end{subfigure}
\caption{Double-slit diffraction.}
\label{fig:double-slit}
\end{figure}

The SCA of Figure \ref{fig:double-slit} describes a vertical screen with two slits, each 20 cells wide (two wavelengths),
separated by 40 cells (four wavelengths). 
The SCA evolves for well over 3000 cycles,
showing interference patterns at both sides of the screen.

\begin{figure}[!h]
  \begin{center}
    \includegraphics[width=0.95\textwidth]{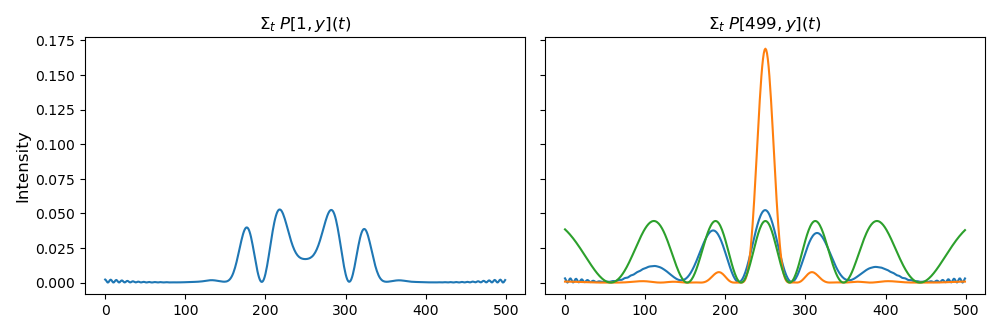}
  \end{center}
  \vspace{-5mm}
\caption{The observed double-slit diffraction pattern (blue) compared to those of 
the Fraunhofer diffraction equation for narrow slits (green) and for wide slits (orange).}
\label{fig:double-slit-diffraction}
\end{figure}

Figure \ref{fig:double-slit-diffraction} compares the observed diffraction pattern
with those described by the Fraunhofer diffraction equation \cite{2025-wiki-Fraunhofer}
for both narrow slits and wide slits:
\begin{equation} 
  I_{narrow}(\phi) ~\propto~ 
  	\cos^2 \Big( \frac{\pi d \sin \phi}\lambda \Big) ~,
 \hspace{1cm}
 I_{wide}(\phi) ~\propto~ 
 	\cos^2 \Big( \frac{\pi d \sin \phi}\lambda \Big)
         \;\mathrm{sinc} \Big( \frac{\pi b \sin \phi}\lambda \Big) ~,
  \label{eq:double-slit-Fraunhofer}
\end{equation} 
where $d$ is the slit separation, $b$ the slit width, and $\phi$ the angle 
with respect to the mid-$y$ horizontal. 
These intensities are normalized across the three central peaks.
Apparently, the observed pattern with $b=2\lambda$ lies somewhere in between.
Note that the observed diffraction pattern stems from a single wave packet,
whereas in a practical double-slit experiment a diffraction pattern
emerges only after the detection of many independent particles.

The double-split experiments of Oliveira et al \cite{2007-Oliveira} describe quantum walks for various QW coins. 
Their Hadamard interference patterns shows a remarkable contrast. 
The interference pattern on the right-hand side of the screen is similar to that of Young's original experiment 
and to that of Figures \ref{fig:double-slit} and \ref{fig:double-slit-diffraction}-(right). 
The pattern on the left-hand side of the screen is distinctly unphysical, 
as if the probabilities tend to stick to the perimeter and the corners of the lattice. 
(No explanation is offered for this.)

\begin{figure}[!h]
   \begin{center}
    \includegraphics[width=0.95\textwidth]{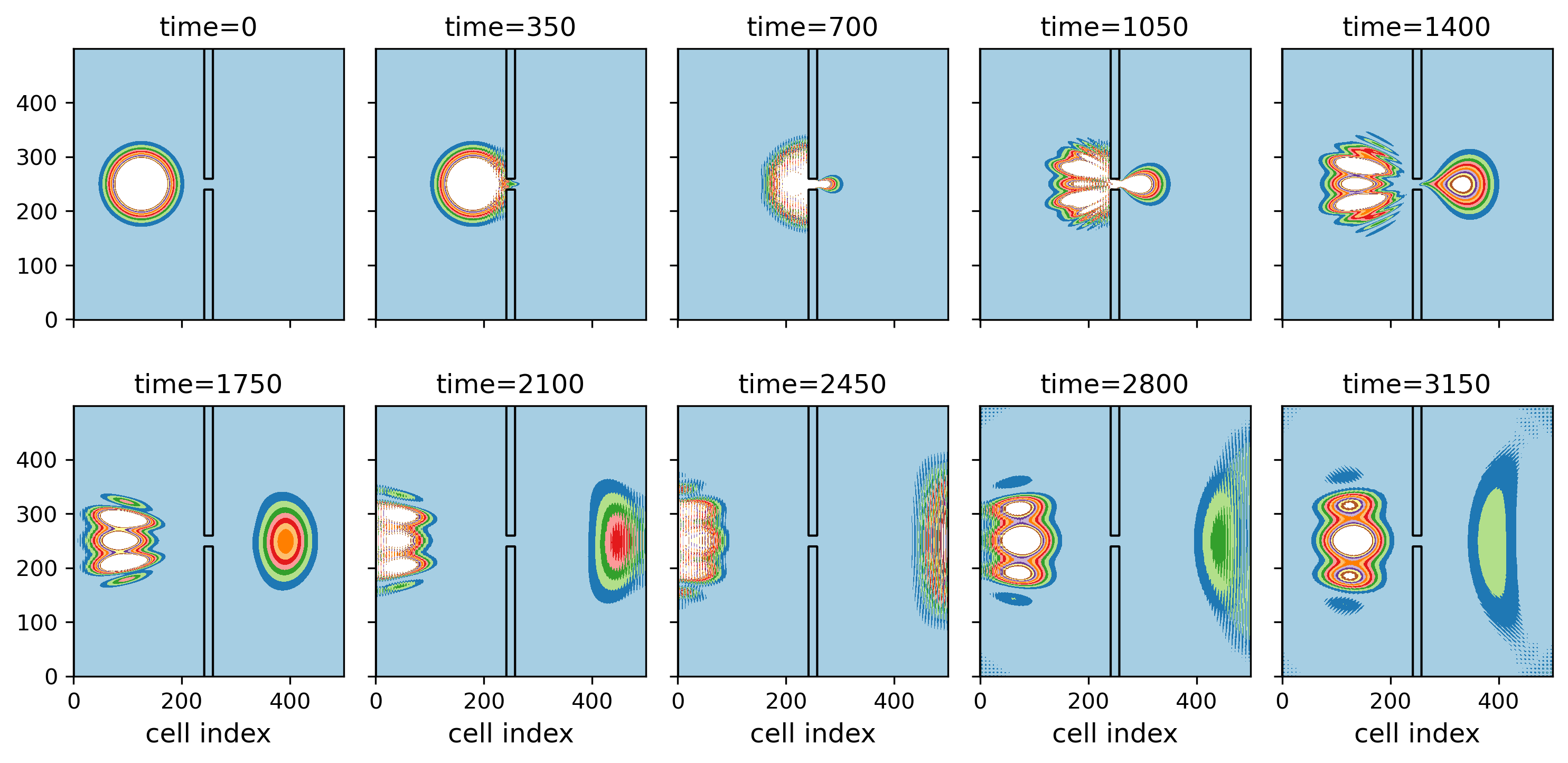}
  \end{center}
  \vspace{-5mm}
\caption{Single-slit diffraction:
      the reflections from the two screen halves interfere on the left-hand side of the screen.}
\label{fig:single-slit-abs}
\end{figure}

The SCA of Figure \ref{fig:single-slit-abs} describes a vertical screen with a single slit. 
Its evolution over about 3000 cycles shows an interference pattern only at the left-hand side
of that screen, resulting from the reflections of the two screen halves.


\subsection{The Aharonov-Bohm effect}
\label{subsec:Aharonov-Bohm}

The Aharonov–Bohm effect is a quantum-mechanical effect 
in which an electrically charged particle is affected by magnetic {\em potential} $\AAA$, 
despite being confined to a region in which magnetic {\em field} $\mathbf{B}$ is zero.
The effect was predicted by Aharonov and Bohm \cite{1959:Aharonov-Bohm}
and experimentally confirmed by Tonomura et al \cite{1982-Tonomura}.
A good introduction is \cite{2009-Feynman-Vol3} Vol.\ II, 15-5.

\begin{figure}[!h]
     \centering
     \begin{subfigure}[b]{\textwidth}
         \centering
         \includegraphics[width=0.95\textwidth]{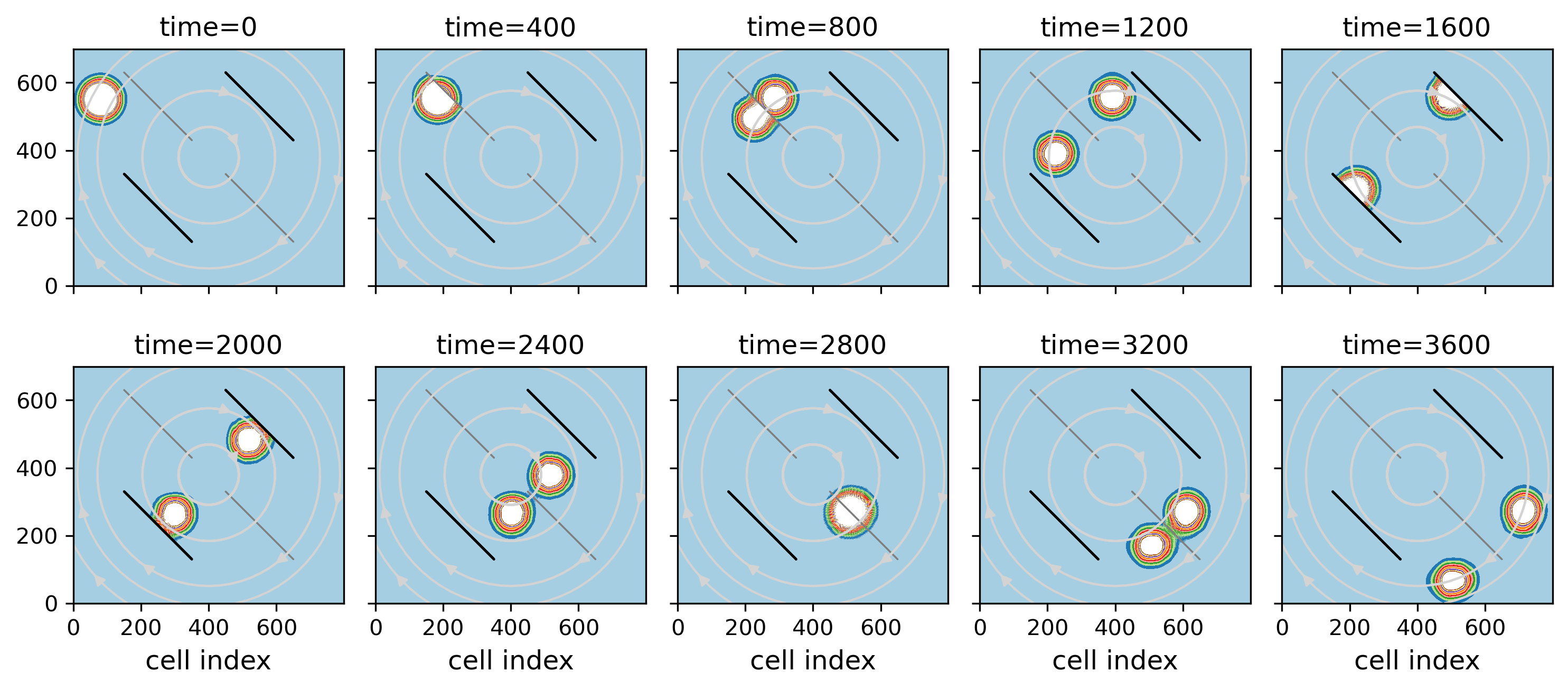}
         \caption{A solenoid is placed at the center of a Mach-Zehnder interferometer.}
         \label{fig:Aharonov-Bohm-MZ}
     \end{subfigure}
     \vspace{3mm}
     
     \begin{subfigure}[b]{\textwidth}
         \centering
         \includegraphics[width=0.95\textwidth]{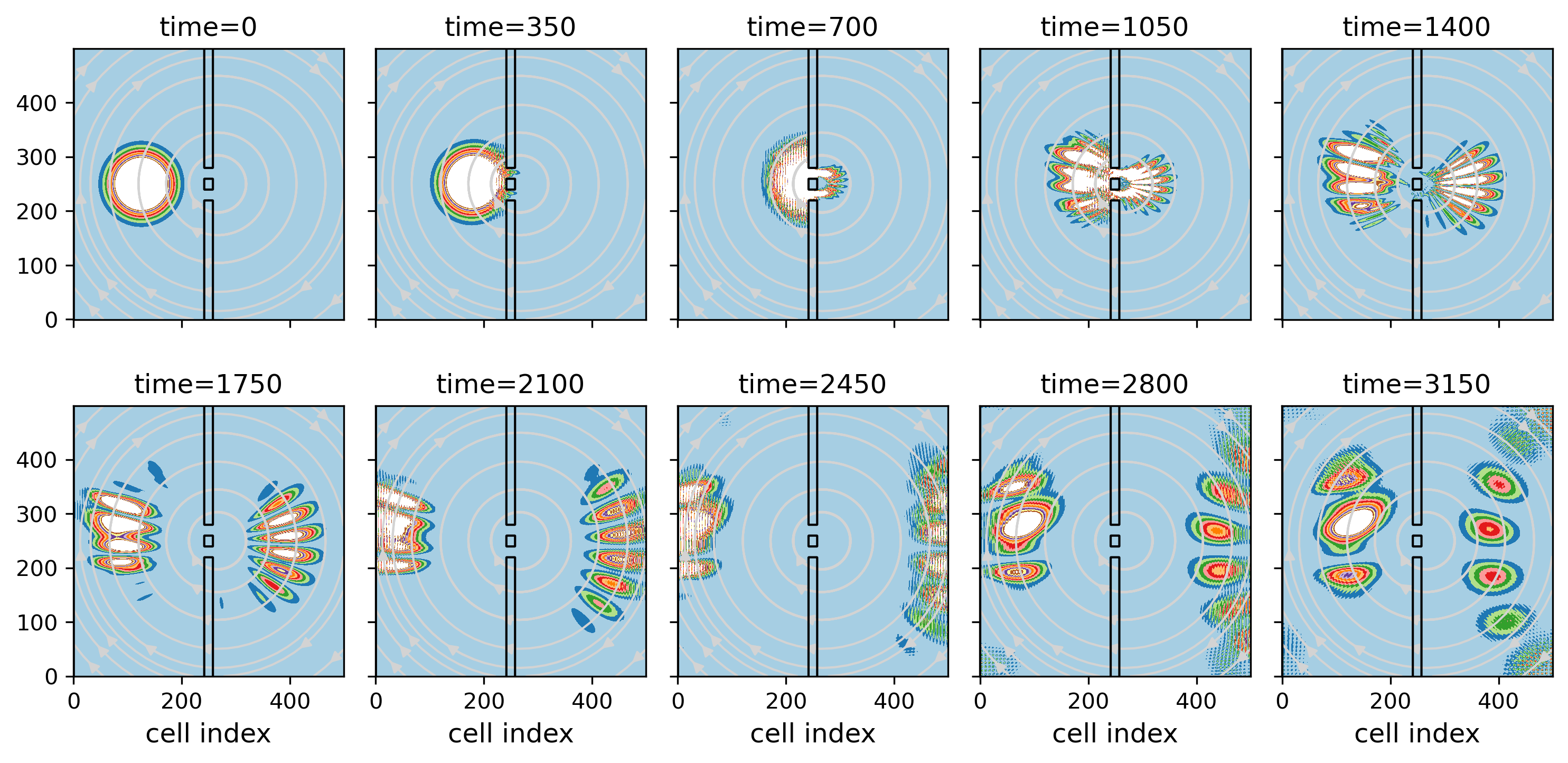}
         \caption{A solenoid is placed just behind the screen center of a double-slit experiment.}
         \label{fig:Aharonov-Bohm-ds}
     \end{subfigure}
     \caption{The Aharonov-Bohm effect. The magnetic potential is shown by gray arrows.}
\end{figure}

\begin{figure}[!h]
  \begin{center}
    \includegraphics[width=0.95\textwidth]{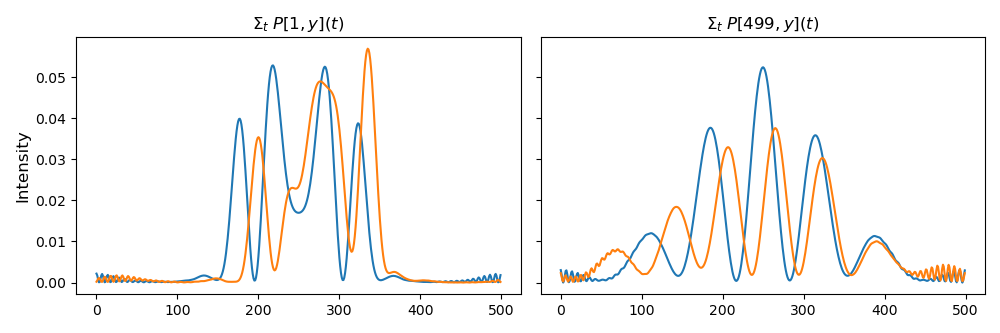}
  \end{center}
  \vspace{-5mm}
\caption{The blue curves show the original double-slit diffraction patterns of Figure \ref{fig:double-slit-diffraction}. 
	The orange curves show a shift to the right, due to the magnetic potential.}
\label{fig:double-slit-diffraction-AB}
\end{figure}

The experiment of Figure \ref{fig:Aharonov-Bohm-MZ} is a variation of that of Figure \ref{fig:Mach-Zehnder-symmetric}, 
where a long solenoid is introduced at the center of a Mach-Zehnder interferometer,
perpendicular to the plane. The magnetic potential of the solenoid is given by 
\cite{2009-Feynman-Vol3}, Vol.\ II, 14-4
\begin{equation} 
	\AAA_{x'} = -K\frac{y'}{x'^2 +y'^2}   
	\hspace{15mm}
	\AAA_{y'} =  K\frac{x'}{x'^2 +y'^2}    
	\hspace{15mm}
	\AAA_{z} =  0  ~,
  	\label{eq:solenoid}
\end{equation}
where $K$ is some constant, and $x'$ and $y'$ are taken relative to the center of the solenoid.
This potential is included in the evolution matrices  $\mathbf{U}_0$ and $\mathbf{U}_1$ 
according to the split Hamiltonian of Subsection \ref{subsec:3D-potentials}.
The presence of this vector potential leads to a different experimental outcome (exit direction of the wave packet).

The experiment of Figure \ref{fig:Aharonov-Bohm-ds} is a variation of that of Figure \ref{fig:double-slit},
where the solenoid is introduced just behind the screen, between the two slits.
Figure \ref{fig:double-slit-diffraction-AB} shows that 
the presence of this potential leads to a right-shift of the diffraction patterns.


\section{Conclusion}
\label{sec:conclusion}

Schr{\" o}dinger cellular automata can be derived from the Schr{\" o}dinger equation 
as follows.
\begin{enumerate}
\item Discretize the continuous Schr{\" o}dinger Hamiltonian and express it matrix form, $\HH$.
\item Split the Hamiltonian matrix $\HH$ (recursively) into a sum of block-diagonal matrices,
         and express it in terms of Hermitian $2 \!\times\! 2$ matrices $\BB_W(\xvec)$ for $W \in \{ X,Y,Z\}$.
\item Approximate $\exp(-i \tau/\hbar\; \HH)$ {\em term by term} (recursively), 
	and express the resulting evolution matrix $\UU$ in terms of
	unitary $2 \!\times\! 2$ matrices $\CB_W(\xvec) = \exp(-i \tau/\hbar \;\BB_W(\xvec)).$
\end{enumerate}
For the free particle this recipe leads to known block (Margolus) unitary cellular automata
(Subsections \ref{subsec:split-hamiltonian} and \ref{subsec:3D-potentials}),
that are mathematically equivalent to a known class of staggered quantum walks.
The recipe naturally extends to 3D inhomogeneous Hamiltonians
(Subsection \ref{subsec:3D-potentials}), 
with scalar potential $V(\xvec)$ and vector potential $\AAA(\xvec)$.
Below the utility of SCAs is assessed from two perspectives: 
1) as a tool for practical single-particle experiments and
2) as a formalism to address the ``what if'' question cited in the introduction.

As tool for analyzing and visualizing non-relativistic single-particle experiments,
it is highly versatile, accurate, and computationally very efficient.
SCA provide:
\begin{itemize}
\item full control over boundaries $X(\xvec)$, scalar potential $V(\xvec)$, vector potential $\AAA(\xvec)$,
and the initial state, all at cell granularity, with $\xvec$ a position in 1, 2, or 3 dimensions,
\item the option to evolve $\UU$ over time (a time-dependent Hamiltonian), and 
\item full observability at cell granularity at every time step.
\end{itemize}
All of these have been demonstrated by a rich set of experiments in Sections 
\ref{sec:1D-experiments} and \ref{sec:2D-experiments}.
The  (systematic) approximation error in the involved physical quantities is $\mathcal{O}\left(\theta^{3}\right)$,
and typically amounts to about $1\%$.
Finally, the intrinsic sparsity of the evolution matrices $\mathbf{U}_0 \mathbf{U}_1$
and $\mathbf{U}_1 \mathbf{U}_0$ results in a very computationally efficient evolutions.
The experiments of this paper typically take only a few minutes on a MacBook.

The introduction highlighted the ``what if'' question:
{\em what if quantum dynamics occurs on a discrete lattice and in discrete time steps, 
e.g. at the Planck scale} \cite{1990-thooft, 2014-thooft, 2014-Arrighi,  2015-bisio}?
D'Ariano coined the phrase {\em Planck cell} \cite{2011-dAriano},
where the cell size $a$ and time step $\tau$  correspond to
the Planck length and the Planck time, with $c = a/\tau$.
A wave packet with a wavelength $\lambda=2a$ 
would have an energy $h\nu = \frac12 h c/a \approx 4\cdot 10^{28} eV$.
This value is many orders of magnitude beyond the 13.6 TeV of the LHC 
and the $\approx 3\cdot 10^{20} eV$ of the most energetic cosmic-ray particles observed.
This suggests that for ``realistic'' wave packets 
the wavelengths must exceed many thousands of cell sizes, 
outside the scope of the experiments of this study.

The most remarkable consequence of the discretization of space
is that velocities become periodic in wavenumber $k$ and energy levels similarly,
as shown in various manifestations in Section \ref{sec:1D-experiments}.
Likewise, evolution matrix $\UU$ becomes periodic in scalar potential $V(\xvec)$, Subsection \ref{subsec:X}.
This unavoidably connects quantum physics to the realm of Nyquist rates and spatial aliasing.
These periodicity properties are not particular to SCA, 
but cellular automata offer an effective and efficient tool to analyze them.
Further exploration of the ``what if'' question, towards relativistic velocities,
suggests additional analysis and experiments with Dirac cellular automata  \cite{2015-bisio}.

\bibliographystyle{plainnat}
\bibliography{Schroedinger_CA} 

\end{document}